\documentclass[twocolumn]{aastex62}
\usepackage{amsmath}
\usepackage[normalem]{ulem}

\graphicspath{{./}{figures/}}
\newcommand{\angstrom}{\textup{\AA}}

\begin{document}

\title{Modeling Dust Production, Growth, and Destruction in Reionization-Era Galaxies with the CROC Simulations II: Predicting the Dust Content of High-Redshift Galaxies}

\correspondingauthor{Clarke Esmerian}
\email{clarke.esmerian@gmail.com}

\author{Clarke J.\ Esmerian}
\affiliation{Department of Astronomy \& Astrophysics, University of Chicago\\
Chicago, IL 60637 USA}
\affiliation{Kavli Institute for Cosmological Physics;
The University of Chicago;
Chicago, IL 60637 USA}
\affiliation{Fermi National Accelerator Laboratory;
Batavia, IL 60510, USA}

\author{Nickolay Y.\ Gnedin}
\affiliation{Fermi National Accelerator Laboratory;
Batavia, IL 60510, USA}
\affiliation{Kavli Institute for Cosmological Physics;
The University of Chicago;
Chicago, IL 60637 USA}
\affiliation{Department of Astronomy \& Astrophysics, University of Chicago\\
Chicago, IL 60637 USA}

\begin{abstract}

We model the interstellar dust content of the reionization era with a suite of cosmological, fluid-dynamical simulations of galaxies with stellar masses ranging from $\sim 10^5 - 10^9 M_{\odot}$ in the first $1.2$ billion years of the universe. We use a post-processing method that accounts for dust creation and destruction processes, allowing us to systematically vary the parameters of these processes to test whether dust-dependent observable quantities of galaxies at these epochs could be useful for placing constraints on dust physics. We then forward model observable properties of these galaxies to compare to existing data. We find that we are unable to simultaneously match existing observational constraints with any one set of model parameters. Specifically, the models which predict the largest dust masses $D/Z \gtrsim 0.1$ at $z = 5$ -- because of high assumed production yields and/or efficient growth via accretion in the interstellar medium -- are preferred by constraints on total dust mass and infrared luminosities, but these models produce far too much extinction in the ultraviolet, preventing them from matching observations of $\beta_{\rm UV}$.  To investigate this discrepancy, we analyze the relative spatial distribution of stars and dust as probed by infrared (IR) and ultraviolet (UV) emission, which appear to exhibit overly symmetric morphologies compared to existing data, likely due to the limitations of the stellar feedback model used in the simulations. Our results indicate that the observable properties of the dust distribution in high redshift galaxies are a particularly strong test of stellar feedback.

\end{abstract}

\keywords{dust -- galaxies: formation -- cosmology: theory -- methods: numerical}

\section{Introduction}

The successful launch and commissioning of JWST has begun a new era in astrophysics. Its unprecedented sensitivity to the emission of high redshift ($z\gtrsim 5$) galaxies has already enabled the rapid accumulation of data on the earliest galaxies. The amount and properties of interstellar dust in these galaxies has a fundamental impact on observations across the entire electromagnetic spectrum, and consequently plays a central role in the understanding of this groundbreaking new data. 

Motivated by this, we have developed a model for the evolution of dust in simulated high-redshift galaxies. Described in ~\citet{Esmerian2022pub}, we post-process simulations from the Cosmic Reionization on Computers project \citep[CROC][]{Gnedin2014, GnedinKaurov2014, Gnedin2016} with a model that determines the fraction of heavy elements in the interstellar medium (ISM) locked in solid dust grains accounting for their nucleation in the ejecta of Asymptotic Giant Brach (AGB) stars and Supernova (SN), their growth via accretion of gas-phase metals in the cold molecular ISM, and their destruction via thermal sputtering due to hot gas from supernova remnants (SNRs). Since the rates of each of these processes are very uncertain due to uncertainties in the material properties of dust grains, the mathematical terms describing them are parameterized with uncertainty factors that enable the exploration of the wide range of theoretical possibilities. 

This model is calculated along Lagrangian tracers that sample gas dynamical quantities along pathlines for a representative fraction of the gas in Lagrangian regions of galaxies. This post-processing technique enables the exploration of dust models with a wide range of parameter choices, of course at the expense of some realism because dust effects are not calculated during simulation run-time. In \citet{Esmerian2022pub}, we explored the full parameter space of the model by focusing on a single massive galaxy ($M_{\rm vir} \sim 10^{11}M_{\odot}$, $M_* \sim 10^9M_{\odot}$ at $z=5$) and the interplay of different dust physical processes in a fully dynamic ISM. We found that reasonable parameter choices for the dust model predicted dust contents and dust-sensitive observables broadly consistent with the extant observational constraints. 

For purposes of computational feasibility, this method development was done using the single most-massive galaxy in a 10$h^{-1}$ cMpc cosmological volume. However, because the initial conditions of large-scale structure are Gaussian random fields, galaxies form in dark matter halos with a wide range of masses and formation histories, which we know to fundamentally impact galaxy properties \citet[see][for contemporary constraints]{Behroozi2019}. Theoretical efforts must therefore strive to make predictions for halos that sample the distributions of masses and formation histories as completely as possible, in order to make predictions for the galaxy population in the real universe. 

This modelling is especially urgent given the recent onslaught of data from JWST, coupled with ambitious programs using radio telescope arrays such as ALMA, that are rapidly fleshing out the properties of the high-redshift galaxy population. Some of the most exciting and puzzling results from this recent revolution have implicated cosmic dust in a central role. There are exciting claims of  anomalously bright galaxies and a surprisingly high star formation rate density at $z > 10$ abound \citep[see][and references therein]{Bouwens2023}, although these are dependent on photometric candidate detections without spectroscopic confirmation and therefore subject to possible revision. If confirmed, reconciling these with the mainstream galaxy formation models may present a challenge, and the many uncertainties of dust enrichment in the first galaxies have been invoked as possible explanations \citep{Mirocha202, Mason2023, Ferrara2023}. Galaxies with spectroscopic confirmation rest on surer footing, and thus far all show evidence for little dust attenuation $z \gtrsim 10$ \citep{Roberts-Borsani2022b, ArrabalHaro2023, Bunker2023, Curtis-Lake2023, Tacchella2023, ArrabalHaro2023b}. 

Nonetheless, the reionization epoch is anything but dust-free. ALMA programs REBELS \citep{Bouwens2022_REBELS, Inami2022} and ALPINE \citep{LeFevre2020} and others \citep{Bowler2022} have detected thermal dust continuum emission that firmly establishes significant amounts of dust in at least some galaxies by $z = 5-7$ \citep{Fudamoto2020, Pozzi2021, Algera2023, Barrufet2023}. These observations also hint at complicated dust morphologies with significant spatial displacement from the stellar component \citep{Bowler2022, Inami2022}. As well, \citet{Rodighiero2023} present an analysis of JWST candidate detections that suggest significant dust obscuration at $8 < z< 13$.  Overall, there is convincing evidence for the very rapid build-up of dust during the reionization epoch, especially in the most massive galaxies. Models of galaxy formation will therefore need to account for the physics of dust if they are to satisfactorily explain key observable constraints on cosmic dawn.

With this goal, in this paper we now extend our previous analysis by applying our dust modelling framework to a suite of 10 additional simulated galaxies from the same simulation volume, selected with approximately uniform logarithmic spacing in final halo mass $1.1\times10^9M_{\odot} \le M_{\rm vir} \le 5.0\times10^{11}M_{\odot}$, corresponding to stellar masses $3.7\times 10^5M_{\odot} \le M_{*} \le 1.9\times10^{9}M_{\odot}$, allowing us to assess the dependence of our predicted dust properties on galaxy mass at a given cosmological time. The paucity of dust at cosmic dawn suggested by some observations motivates us to also explore a wider range of dust modelling choices, namely those that either produce less dust or destroy it more efficiently. Section \ref{methods} explains our simulated galaxy sample selection, notes small updates to the methodology presented in the first paper, and presents the dust model variations explored in this analysis. Section \ref{results} presents the galaxy mass-metallicity relation predicted by the simulations compared to existing high-redshift constraints, and results of the dust model applied to our simulated galaxy sample. Specifically, we present the predicted dust content and dust-sensitive observable quantities, both galaxy-averaged and spatially resolved, to which we compare to existing data. Section \ref{discussion} discusses the agreements and discrepancies between our model predictions and observational constraints, and compares our work to other recent similar investigations in the literature. We conclude in Section \ref{conclusion}.

\section{Methods}

\label{methods}
The galaxy formation simulation model, halo identification, and galaxy definitions are identical to those described in \citet{Esmerian2022pub}, to which we refer the reader. For this paper's analysis, we select a total of 11 galaxies from a 10$h^{-1}$ co-moving Megaparsec (cmMpc) cosmological volume with final $z=5$ halo masses  $1.1\times10^9M_{\odot} \le M_{\rm vir} \le 5.0\times10^{11}M_{\odot}$, corresponding to final stellar masses $3.7\times 10^5M_{\odot} \le M_{*} \le 1.9\times10^{9}M_{\odot}$. These limits span the range of halo masses resolved in the simulation. The 11 halos are selected with approximately logarithmically uniform spacing in final halo mass. Since the galaxy scaling relations predicted by CROC have small scatter (see \citet{Zhu2020} and \citet{Noel2022}) and do not dramatically change slope on scales $\lesssim 0.5$ dex in halo mass, it is sufficient for the purposes of this analysis to sample one halo of a given mass with the average spacing of $0.24$ dex provided by a total sample of 11. This simulation has the same initial conditions as the one used in the previous paper, so the most massive halo is the same.

As in the previous paper, we sample ISM conditions in these simulated galaxies along pathlines traced by Lagrangian tracer particles. These particles are initialized in random positions in the Lagrangian region of the halo and follow the fluid flow using the Monte-Carlo method introduced in \citet{Genel2013} and implemented in the ART code in \citet{Semenov2018}. The number of tracers per halo scales with halo mass such that the minimum number of tracers for a given halo is above 100. For galaxies hosted in halos with final masses $M_{\rm vir} > 10^{11}M_{\odot}$, we downsample to $10^4$ particles for computational feasibility. 

\begin{figure}
    \centering
    \includegraphics[width=\linewidth]{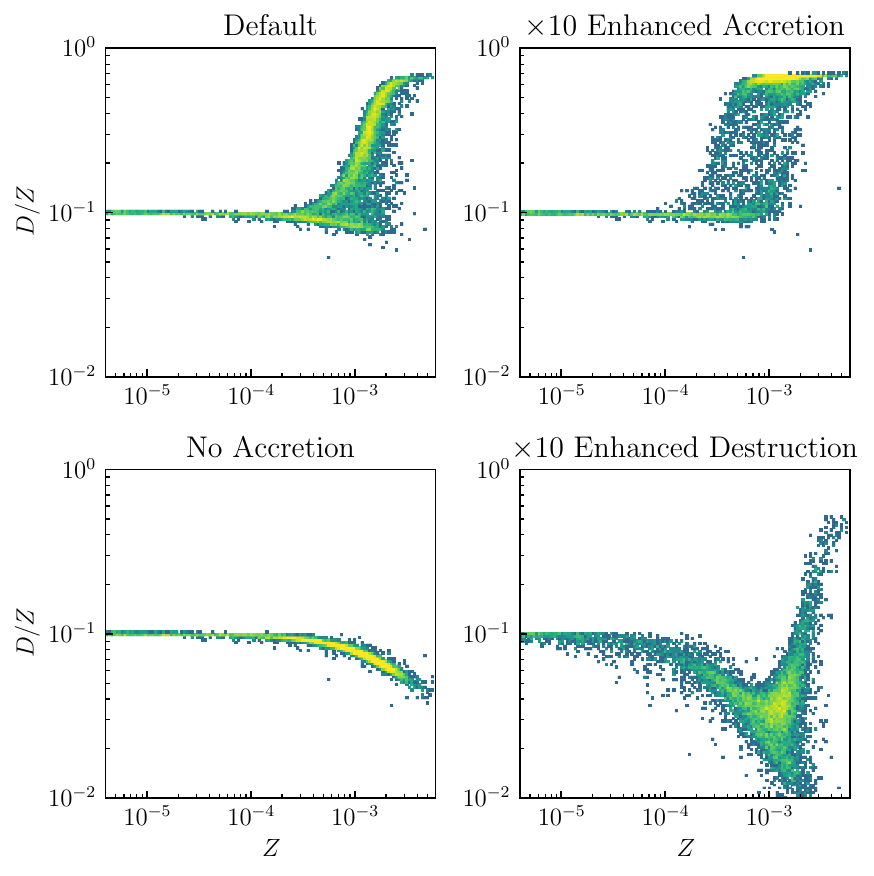}
    \caption{Spatially-resolved $D/Z$ vs $Z$ relations. Each panel shows the 2D PDF of mass for $D/Z$ vs $Z$ in the ISM of the most massive galaxy at the last snapshot. Different panels correspond to different dust models.}
    \label{fig:DtoZ_corr_last}
\end{figure}

Computational resource limitations therefore prevent us from using enough tracers that all cells in each galaxy are sampled. Consequently, we must find a way to assign dust masses to cells in the galaxy which were not sampled by any tracer. To do this, we interpolate the $D/Z$ vs $Z$ relation for the tracers in each galaxy at every snapshot output. As shown in Figure \ref{fig:DtoZ_corr_last}, the dust-to-metal ratio scales regularly with metallicity, making this interpolation the best option for assigning dust masses to unsampled cells in a way that preserves the predictions of the dust model. We note that with some model choices even this relation exhibits significant scatter at a given tracer metallicity. Our correction therefore possibly underestimates the scatter in observable quantities impacted by the dust distribution.

We note that in \citet{Esmerian2022pub} we had not yet realized this correction was necessary, and therefore the results in that paper underestimate the total dust mass and effect of dust on observable quantities. These effects are $\mathcal{O}(10\%)$ in the dust mass but can be order-unity for some observable quantities that depend sensitively on the dust distribution -- particularly dust extinction of UV starlight -- but do not change the qualitative conclusions of that paper.

Otherwise, the methods used to calculate dust-dependant observables (the effective optical depth to dust at $1500\angstrom$: $\tau_{\rm 1500}$, the logarithmic spectral slope in the UV: $\beta_{\rm UV}$, the infrared luminosity: $L_{\rm IR}$ and the infrared excess IRX$\equiv L_{\rm IR}/L_{\rm UV}$) are identical to the description in Section 2.6 of \citet{Esmerian2022pub}. 

 However, we note that the published $L_{\rm IR}$ values in that paper were incorrectly calculated by integrating the dust infrared emission from 2.5$\AA$ to 20$\AA$, instead of $8\AA$ to $1000\AA$ as stated in the text, due to a bug in the relevant code. Consequently, the infrared luminosities in that paper are incorrect by a factor of $\sim 10^3$, and we show the effect of this correction in Appendix~\ref{appendix:L_IR}. However, the scaling of infrared luminosities with dust mass predicted by our analysis was nonetheless qualitatively correct, and since we did not make any comparison to observations for these quantities, the main conclusions of that paper remain unchanged. 

 \subsection{Spatially Resolved Images}

For the analysis of spatially resolved dust properties we focus exclusively on the most massive galaxy in the simulation which attains a stellar mass of $M_* = 1.3\times10^8M_{\odot}$ by $z=8$, $M_* = 7.1\times10^8M_{\odot}$ by $z=6.4$ and $M_* = 1.9\times10^9M_{\odot}$ by $z=5$. Our simulated galaxy is therefore well within the range of stellar masses probed by observations to which we compare \citep{Inami2022}, justifying comparison because galaxy morphology is expected to depend on stellar mass \citep{Pillepich2019}. However, we note that for these snapshots it has a much lower SFR of $\le 5.2M_{\odot}/{\rm yr}$ than any galaxy in this sample. This may be due to the bias towards high SFR systems of these observations noted in the introduction, or the inability of the CROC model to produce such rapidly star-forming systems. Figure 6 of \citet{Zhu2020} shows that the CROC galaxies exhibit small scatter in the SFR-stellar mass relationship, suggesting that the discrepancy between our simulation's SFR and those inferred from data results from a deficiency of the model and would not be alleviated by considering a larger number of simulated galaxies. This discrepancy provides further motivation to compare our simulations to data in a spatially resolved analysis as this may provide information about the cause of these discrepancies.

We present results for this simulation at 12 snapshots from $z = 8.5$ to $z=5$. The upper bound in redshift is motivated by the upper bound on the observations, but we include snapshots lower than the observational lower bound of $z=6.5$ to maximize the galaxy mass range probed by our analysis. We note that we have redone the analysis restricted to only snapshots within $6.4 < z < 8.5$, identical to the observations, and all of our conclusions are unchanged. We also note that based on visual inspection, the galaxy undergoes merging events at $z\approx 7.3$ and $z\approx 6$ which significantly disrupt its morphology. There are enough snapshots of the simulation that the galaxy morphology before, during, and after the merger event can be clearly distinguished. We therefore expect that our simulation data samples a sufficiently violent merger history with sufficiently high time resolution that our analysis accounts for the morphological effects of accretion history on high-redshift galaxies of the relevant masses.

This spatial analysis uses quantities calculated as described previously except with small modifications. UV colors are determined based on the finite difference between luminosities at $1500\angstrom$ and $2500\angstrom$ as follows

\begin{equation}
    \beta_{\rm UV} = \frac{\log_{10}(f_{2500\angstrom}/f_{1500\angstrom})}{\log_{10}(2500\angstrom/1500\angstrom)}
\end{equation}

\noindent again on an individual star particle basis. We note that this is not identical to the calculation of $\beta_{\rm UV}$ in the rest of the analysis, in which a least-squares fit performed on this portion of the UV spectrum to determine a power-law slope. This finite difference method is adopted for computational ease, and we have checked that it reproduces the least-squares fitting results very accurately. 

We use dust column density $\Sigma_D$, computed from the galaxy dust mass distribution calculation as a proxy for infrared continuum emission. The two are directly proportional because the dust distribution is optically thin in the infrared. 

To account for the effect of finite observational resolution, we smooth with a Gaussian kernel with variance $\sigma^2 = \frac{\Delta x_{\rm FWHM}^2}{8\ln{2}} $ where $\Delta x_{\rm FWHM}$ is the physical distance at the simulation snapshot redshift corresponding to the angular Full-Width Half-Max (FWHM) of the observation. We explore $\Delta x_{\rm FWHM}$ values of $0.2$ and $0.8$ arcsec, which for the redshift range of our simulations $11.4 \ge z \ge 5.0$ corresponds to a physical size of $0.33-0.55$kpc and $1.32-2.19$kpc, respectively.

\subsection{Dust Model Parameter Exploration}\label{subsec:models}

\begin{deluxetable*}{|c|c|c|c|}
\tabletypesize{\footnotesize}
\tablecaption{Explored Parameter Combinations. Note that for each model, any parameter not listed under Key Parameters is the same as the Default model.\label{table:models}}
\startdata
\tablehead{
    {\bf Model Name} & 
    {\bf Key Parameters} & 
    {\bf Description} & 
    {\bf Color in Figures}
    }
{\bf Default} & $y_{D,{\rm SN}} = y_{D,{\rm AGB}} = 0.1$, & Default from \citet{Esmerian2022pub}, & {\includegraphics[scale=0.1]{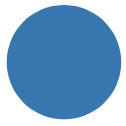}} \\
& $\tau_{\rm accr} = 3\times10^8$yr, $C_{\rm dest} = 1$ &  parameters based on \citet{Dwek1998, Feldmann2015}. & \\
\hline
{\bf No Accretion} & $\tau_{\rm accr} = \infty$ & No grain growth due to gas-phase accretion in cold ISM & {\includegraphics[scale=0.1]{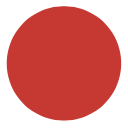}} \\
\hline
{\bf Enhanced Accretion} & $\tau_{\rm accr} = 10\tau_{\rm accr, Default}$ & Enhanced grain growth  &{\includegraphics[scale=0.1]{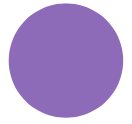}}\\
& & due to gas-phase accretion in cold ISM & \\
\hline
{\bf No Destruction} & $C_{\rm dest} = 0$ & No grain destruction in hot gas due to SNRs & {\includegraphics[scale=0.1]{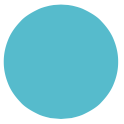}}\\
\hline
{\bf Enhanced Destruction} & $C_{\rm dest} = 10C_{\rm dest, Default}$ & Enhanced grain destruction in hot gas due to SNRs & {\includegraphics[scale=0.1]{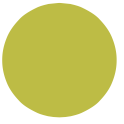}} \\
\hline
{\bf Very Enhanced Destruction} & $C_{\rm dest} = 100C_{\rm dest, Default}$ & Very enhanced grain destruction in hot gas due to SNRs & {\includegraphics[scale=0.1]{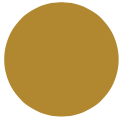}} \\
\hline
{\bf Low SN Production} & $y_{D,{\rm SN}} = 0.1y_{D,{\rm SN, Default}}$ & Suppressed dust yield from SN &  {\includegraphics[scale=0.1]{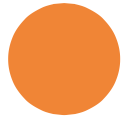}} \\
\hline
{\bf No SN Production} & $y_{D,{\rm SN}} = 0$ & SN do not produce dust &  {\includegraphics[scale=0.1]{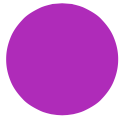}} \\
\hline
{\bf Very Low SN Production,} & $y_{D,{\rm SN}} = 0.1y_{D,{\rm SN, Default}}$,& Very suppressed dust yield from SN, & {\includegraphics[scale=0.1]{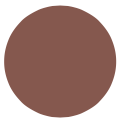}}\\
{\bf  No AGB Production} &  $y_{D,{\rm AGB}} = 0$ &  AGB do not produce dust & \\
\hline
\enddata
\end{deluxetable*}

As in the previous paper, we run a suite of dust models with different parameter choices to explore their impact on the predicted dust content of high-redshift galaxies and dependent observables, now on a sample of multiple simulated galaxies. However, motivated by observations that increasingly point to minimal dust in the earliest galaxies \citep[e.g.][]{Roberts-Borsani2022b, Tacchella2023}, we extended the set of parameter variations explored by introducing 3 new models that either increase the grain destruction rate in supernova remnants ({\bf Very Enhanced Destruction}) or decrease grain production in SN and/or AGB ({\bf No SN Production}, and {\bf Very Low SN Production, No AGB Production}). The list of models explored in this analysis are summarized in Table \ref{table:models} and described below. Note that each model is assigned a unique color for Figures 4-9, and 11 which are shown in the right-most column of the table.

\begin{itemize}
    \item {\bf Default:} This is identical to the ``Default'' model explored in \citet{Esmerian2022pub}, for which parameters were chosen to be the same as successful similar physical models of dust evolution for local-universe galaxies \citep{Dwek1998, Feldmann2015, Li2019}.
    \item {\bf No Accretion:} This is identical to the ``No Accretion'' model in \citet{Esmerian2022pub}. The parameters of this model are identical to {\bf Default} except grain growth due to accretion of gas-phase metals in the cold molecular ISM is not allowed. This parameter choice is motivated by arguments based on microphysical considerations of dust grain geometry that grain growth in the cold phase of the ISM may not be possible \citep{Ferrara2016}.
    \item {\bf Enhanced Accretion:} This is identical to the ``Enhanced Accretion'' model in \citet{Esmerian2022pub}. The parameters of this model are identical to {\bf Default} except grain growth due to accretion of gas-phase metals in the cold molecular ISM is enhanced by an order of magnitude. This parameter choice is motivated both by uncertainties in the unresolved density distribution of the cold ISM in our simulations, where grain growth is expected to be most efficient, and to enable comparison to other works that adopt faster grain growth rates \citep{Graziani2020, Lewis2023}.
    \item {\bf No Destruction:} This is identical to the ``No Destruction'' model in \citet{Esmerian2022pub}. The parameters of this model are identical to {\bf Default} except grain destruction in the hot gas of SNRs is not allowed. This parameter choice is motivated by indirect observational indications of inefficient dust destruction in high-temperature gas \citep{GallHjorth2018, Gjergo2018, Vogelsberger2019, Michalowski2019}, as well as uncertainties in the unresolved ISM phase structure in our simulations.
    \item {\bf Enhanced Destruction:} This is identical to the ``Enhanced Destruction'' model in \citet{Esmerian2022pub}. The parameters of this model are identical to {\bf Default} except grain destruction in the hot gas of SNRs is enhanced by an order of magnitude. This parameter choice is motivated by uncertainties in the destruction efficiency of individual supernova remnants both due to the microphysics of dust and unresolved ISM phase structure \citep{McKee1989, Hu2019, Kirchschlager2022}.
    \item {\bf Very Enhanced Destruction:} The parameters of this model are identical to {\bf Default} except grain destruction in the hot gas of SNRs is enhanced by two orders of magnitude. The motivation for this parameter choice is the same as for {\bf Enhanced Destruction}, since the associated uncertainties are large, and also the increasing evidence for dust-free early galaxies as mentioned previously.
    \item {\bf Low SN Production:} Identical to {\bf Default} except the dust yield from supernova is supressed by an order of magnitude (i.e. $y_{D,{\rm SN}} = 0.01$). Note that we do not change the AGB yield $y_{D,{\rm AGB}}$, and since the AGB metal production is about 10 times smaller than that of SN \citep[see][Figure 7]{Esmerian2022pub}, SN and AGB production are comparable with these parameters. This parameter choice is also motivated by the evidence for minimally dusty high-redshift galaxies, and uncertainties about the fraction of SN-produced dust that survives the reverse shock \citep[see e.g.][]{BianchiSchneider2007, Michelotta2016, Slavin2020}.
    \item {\bf No SN Production:} Identical to {\bf Default} but SN production is turned off -- $y_{D,{\rm SN}} = 0$. This choice is motivated by the extreme scenario in which no dust survives the reverse shock of any supernova.
    \item {\bf Very Low SN Production, No AGB Production:} Identical to {\bf Default} except the dust yield from supernova is suppressed by two orders of magnitude (i.e. $y_{D,{\rm SN}} = 10^{-3}$) and AGB production is turned off ($y_{D,{\rm AGB}} = 0$). This is motivated by the same considerations as for the previous two models, and the deep uncertainties around AGB dust production especially in the early universe \citep[e.g.][]{Valiante2009, Schneider2014, DellAgli2019, Tosi2023}.
\end{itemize}

\section{Results}
\label{results}
\subsection{The Mass-Metallicity Relation}

\begin{figure}
    \centering
    \includegraphics[width=\linewidth]{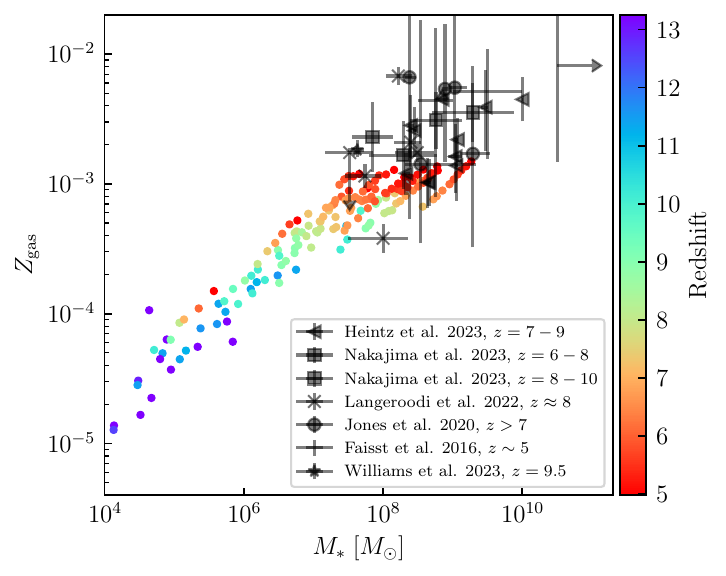}
    \caption{Mass-Metallicity Relation. The galaxy-averaged gas-phase metallicity is shown as a function of stellar mass. Each point represents an individual galaxy at an individual snapshot, colored by redshift. Observational data from galaxies at similar redshifts are from \citet{Faisst2016, Jones2020, Langeroodi2022, Nakajima2023, Heintz2022}, and \citet{Williams2023} converted to mass fraction using 12 + [O/H]$_{\odot}$ = 8.71 and $Z_{\odot} = 0.02$ \citep{Lodders2019}} 
    \label{fig:galaxy_scaling_relations}
\end{figure}

The dust content of galaxies is normalized by their overall metal content, so we first examine the galaxy metallicities in the simulations. Figure ~\ref{fig:galaxy_scaling_relations} shows the mass-metallicity relation for our simulations, including existing data at relevant redshifts. While the data mainly overlap with only the highest-mass galaxies in our sample, where there is overlap we see fair agreement, albeit with some indications of systematically low metallicities in CROC. The galaxy scaling relations of other quantities predicted by CROC have been thoroughly discussed and compared to existing data in \citet{Zhu2020}. We note that \citet{Noel2022} presented a more detailed analysis of the CROC mass-metallicity relation, but at the time these high-redshift data were not available for comparison, making this a new result.

\subsection{The Dust Content of High-Redshift Galaxies}

\begin{figure*}
    \centering
    \includegraphics[width=\linewidth]{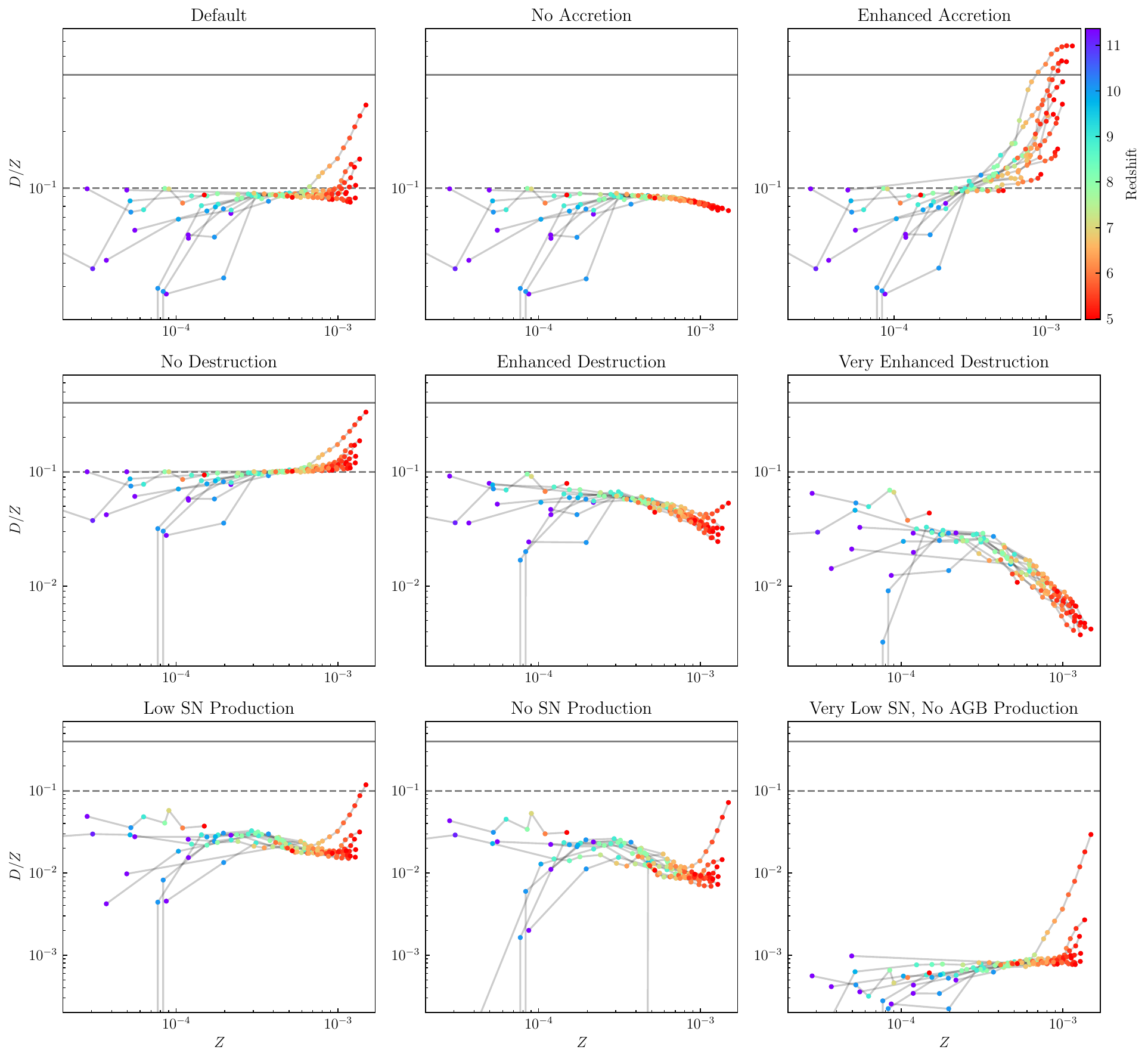}
    \caption{Galaxy-Averaged $D/Z$ vs. $Z$ relations. Each panel shows the evolution of the dust content in our simulated galaxies with a different set of assumed dust model parameters, as indicated by the titles. Each point corresponds to the average $D/Z$ and $Z$ values for an individual galaxy at an individual snapshot. $Z$ is in physical (i.e. mass fraction) units. Points from the same galaxy are connected with grey lines, and colors indicate redshift. The dashed horizontal grey line indicates $D/Z = 0.1$, which is the default production yield in our model and the solid horizontal grey line indicates $D/Z = 0.4$, the value for the Milky Way and a common choice in post-processing analyses (see Section \ref{subsec:discussion_comparison}).}
    \label{fig:DtoZ_Z}
\end{figure*}

\begin{figure*}
    \centering
    \includegraphics[width=\textwidth]{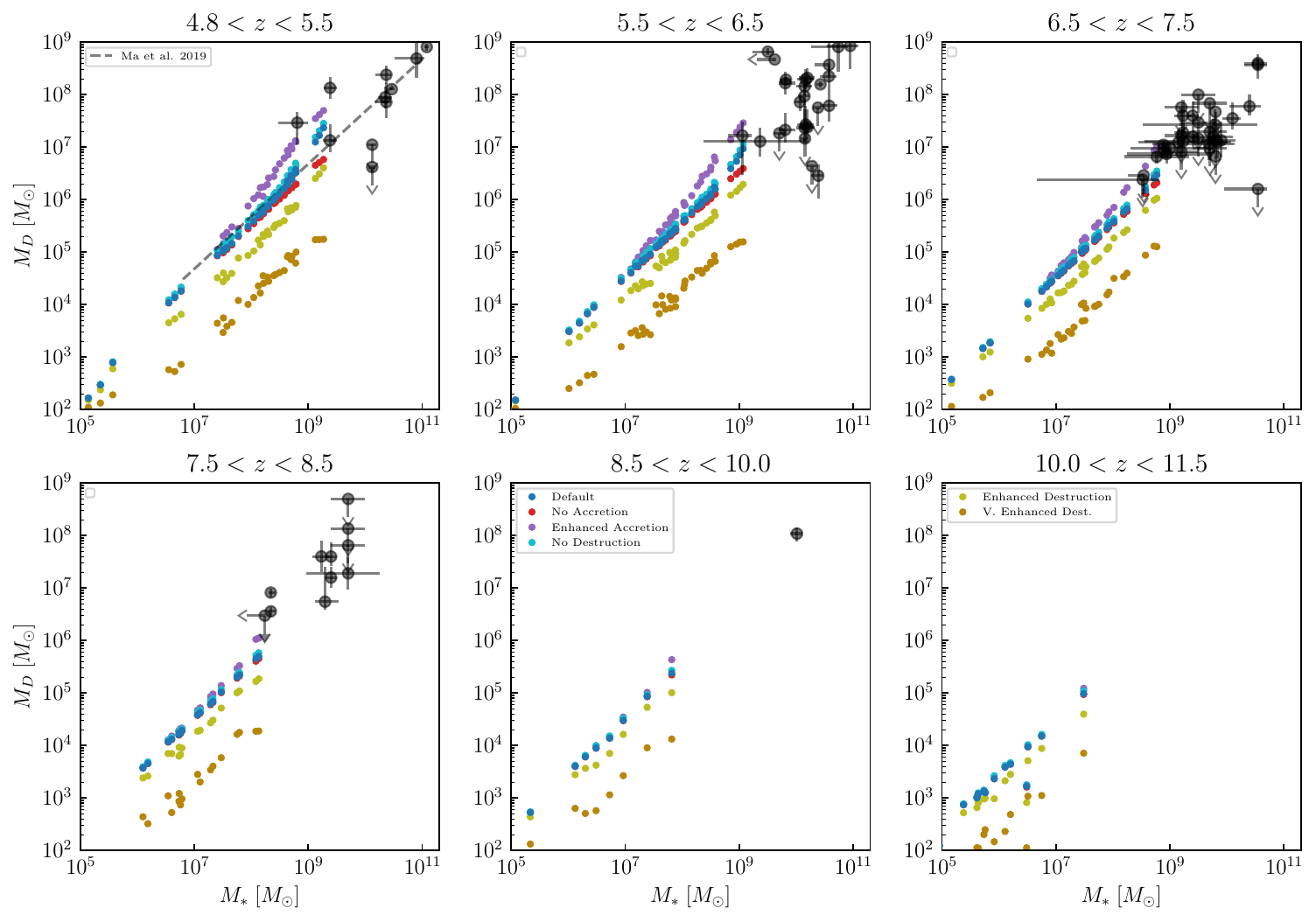}
    \caption{Dust mass-stellar mass relation. Colors indicate different dust models which include {\bf Default} and those with varied growth or destruction parameters (i.e. the first two rows of those shown in Fig.~\ref{fig:DtoZ_Z}). Each point is a single galaxy at a single redshift, and separate panels are redshift bins. Estimates based on observational data from \citet{Sommovigo2022}, \citet{Dayal2022}, \citet{Hashimoto2019}, \citet{Knudsen2017}, \citet{Schaerer2015}, \citet{Watson2015}, \citet{Laporte2017}, \citet{Tamura2019}, \citet{daCunha2015}, \citet{Marrone2018}, \citet[][with a redshift for ID27 from \citet{Aravena2016}]{Burgarella2020}, \citet[][with stellar masses from \citet{Faisst2020}]{Pozzi2021}, \citet{Witstok2023},  and \citet{LesniewskaMichalowski2019} are shown with the same redshift binning. The predictions of a simpler dust post-processing model on higher-resolution simulations presented in \citet{Ma2019} is shown in the dashed grey line.}
    \label{fig:MD_Mstar_corr_acc_dest}
\end{figure*}

\begin{figure*}
    \centering
    \includegraphics[width=\textwidth]{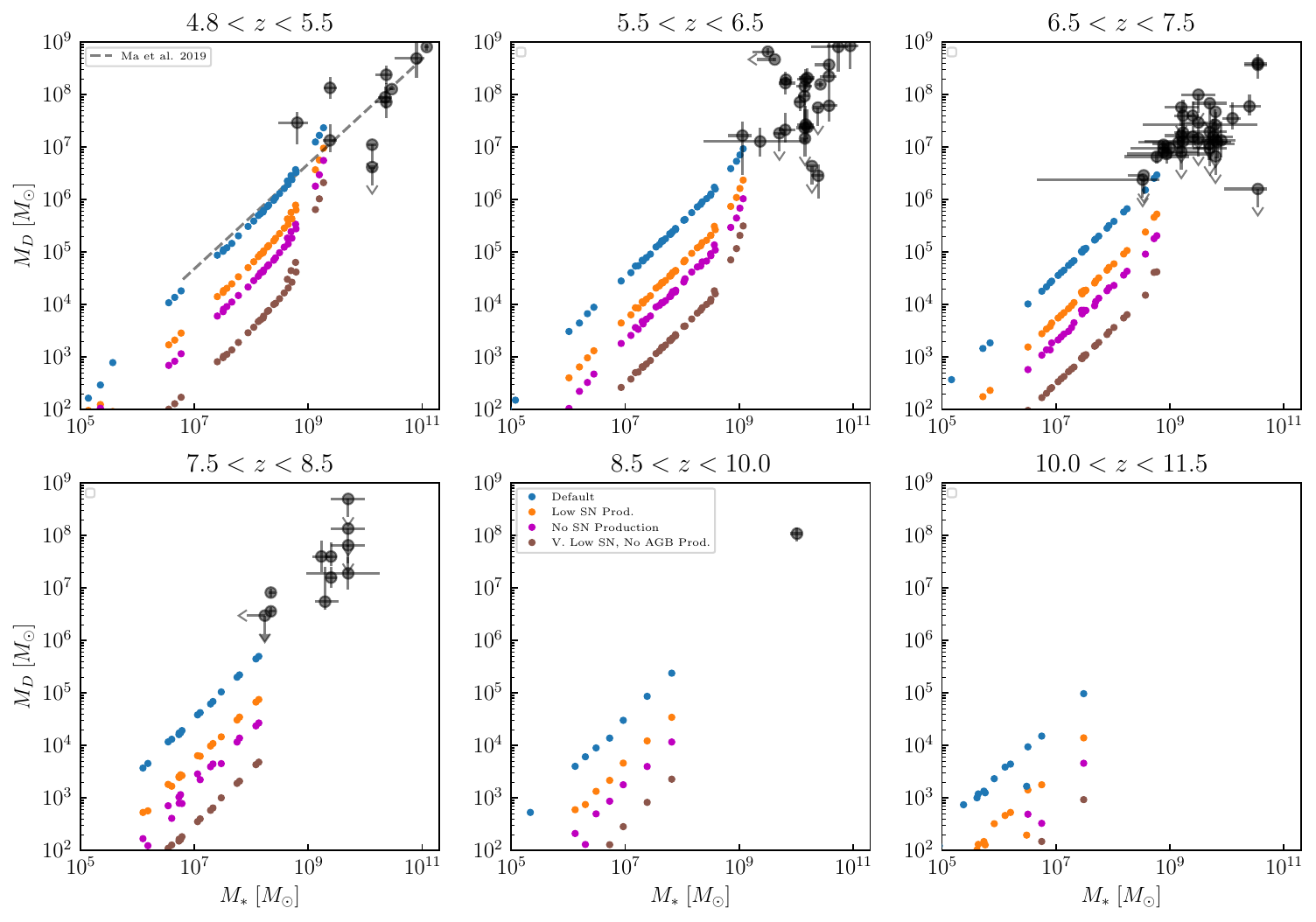}
    \caption{Dust-mass stellar-mass relation cont'd. Same as Fig.~\ref{fig:MD_Mstar_corr_acc_dest} but with {\bf Default} and the dust models with varied yields (i.e. those in the third row shown in Fig.~\ref{fig:DtoZ_Z}).}
    \label{fig:MD_Mstar_corr_yD}
\end{figure*}

The dust content of our simulated galaxies predicted by the models described in Section \ref{subsec:models} and Table \ref{table:models} are summarized in Fig.~\ref{fig:DtoZ_Z}, where we show the galaxy-averaged dust-to-metal ratio $D/Z$ as a function of galaxy metallicity $Z$ in mass-fraction units (i.e.\ in which the average metallicity in the solar neighborhood is $Z_\odot=0.02$). The top row shows the default model and variations in the ISM grain growth accretion timescale. The middle row shows variations in the grain destruction efficiency of supernova remnants. And the bottom row shows variations in the assumed yields of dust production sources.

Broadly, we notice several trends. A well-established property of dust models similar to ours is the transition of dominant physical processes between low and high-metallicity regimes: the $D/Z$ ratio at low metallicity ($Z\lesssim 4\times10^{-4}=2\times10^{-2}Z_{\odot}$) is primarily set by the choice of source yields $y_{D,{\rm SN/AGB}}$, while the ratio at high metallicity is determined by a competition between the timescale for grain growth due to accretion in the ISM, and the efficiency of grain destruction in the hot ionized medium. If grain growth dominates, (as in {\bf Default}, {\bf Enhanced Accretion}, {\bf No Destruction}, {\bf Low SN Production}, {\bf No SN Production}, and {\bf Very Low SN, No AGB}) the $D/Z$ ratio rises with increasing metallicity, while when destruction dominates ({\bf No Accretion}, {\bf Enhanced Destruction}, {\bf Very Enhanced Destruction}) the opposite scaling is observed.

However, we note that in some models there is substantial scatter between galaxies even at fixed metallicity, particularly for those models where accretion dominates at late times. The efficiency of grain growth via accretion must therefore depend on galaxy properties beyond average metallicity, an effect not captured in simpler one-zone models \citep[e.g.][]{Feldmann2015}. For the {\bf Default} and {\bf Enhanced Accretion} models we notice substantial scatter in the metallicity at which each galaxy enters the growth-dominated regime of rising $D/Z$. 

Indeed, we note that the most massive galaxy in our sample exhibits rising $D/Z$ at high metallicities in all models except {\bf No Accretion} (where $D/Z > y_D$ is physically impossible) and {\bf Very Enhanced Destruction}. Even in the {\bf Enhanced Destruction} scenario the $D/Z$ ratio rises at late times (i.e. high metallicies) for this single galaxy but no others. This clearly indicates the importance of some combination of star formation history and ISM phase structure in setting the dominant dust regulating mechanisms. Precise determination of this cause would require more analysis beyond the scope of this work but would be interesting for future investigation.

Finally, we note that the significant scatter in $D/Z$ at very low metlallicities $Z \lesssim 10^{-4}$. As indicated by their redshift (indicated in color) and the stellar mass-metallicity relation in Figure~\ref{fig:galaxy_scaling_relations}, these low-metallicity galaxies are the highest-redshift and lowest-mass in our sample. Consequently, there are the most poorly-resolved and subject to the greatest stochasticity effects from the discreteness of enrichment from star particles and sampling by Lagrangian tracers. The latter would be amended by coupling the dust model explicitly to the simulation, and is therefore another motivation for more sophisticated modelling in future analyses. Nonetheless, this noise occurs at such low metallicities that its effect on the total dust mass, which is normalized by the metallicity, is minor and should not strongly impact our conclusions. As well, the existence of clear trends at late times/high metallicities indicates that the predictions are well resolved for the most massive galaxies, which are the most relevant for comparison to observational data.

In Figures \ref{fig:MD_Mstar_corr_acc_dest} and \ref{fig:MD_Mstar_corr_yD} we examine the predicted dust masses of our simulated galaxies as a function of stellar mass at different redshifts for the different dust models. Figure \ref{fig:MD_Mstar_corr_acc_dest} shows {\bf Default} and models with variations in accretion and destruction rates (the first and second rows of Figure~\ref{fig:DtoZ_Z}), while Figure \ref{fig:MD_Mstar_corr_yD} shows models with varied production yields (third row of Figure~\ref{fig:DtoZ_Z}. In all cases, the dust mass exhibits an approximately linear scaling with stellar mass, with varied normalization depending on assumed production yields and destruction efficiencies. This normalization spans two dex at a given stellar mass for the entire suite of models herein considered. There is also a general steepening of the relationship at higher masses ($M_* \sim 10^8-10^9M_{\odot}$) in models where accretion becomes efficient. These relationships are sufficiently tight to be well-distinguished between different models in principle, although there is significant degeneracy between yield and destruction rates -- {\bf Very Enhanced Destruction} and {\bf No SN Production} predict very similar values which the first achieves by destroying dust with high efficiency while the second produces little dust to begin with. As well, models with the same yield but different growth timescales ({\bf Default}, {\bf No Accretion}, {\bf Enhanced Accretion}) are only distinguishable at high masses and late times, consistent with the results of Figure ~\ref{fig:DtoZ_Z}. In summary, different plausible parameter choices for the dust model can change dust masses by up to two orders of magnitude at a given stellar mass. This flexibility highlights the need for a significant improvement in our understanding of dust production and destruction processes.

We therefore compare these predictions with existing observational estimates of dust masses in high-redshift galaxies from \citet{Sommovigo2022}, \citet{Dayal2022}, \citet{Hashimoto2019}, \citet{Knudsen2017}, \citet{Schaerer2015}, \citet{Watson2015}, \citet{Laporte2017}, \citet{Tamura2019}, \citet{daCunha2015}, \citet{Marrone2018}, \citet[][with a redshift for ID27 from \citet{Aravena2016}]{Burgarella2020}, \citet[][with stellar masses from \citet{Faisst2020}]{Pozzi2021}, \citet{Witstok2023},  and \citet{LesniewskaMichalowski2019}. Because of the limited volume of our simulation, we do not capture unusually massive and therefore rare halos, limiting us to predictions at lower masses than almost all the existing data. Nonetheless, the data appear to favor those models with the highest dust masses -- the data is always at the upper envelope of our simulation predictions wherever they overlap. Indeed, in both the $6.5 < z< 7.5$ and $7.5 < z < 8.5$ bins most of the data appear to lie on or above the scaling relation of the most dust-rich model {\bf Enhanced Accretion} if it were extrapolated. This suggests that the data prefer models in which production yields are high and ISM grain growth is efficient at high masses. We also note that the data appear to exhibit greater scatter at a given stellar mass than any one set of dust model parameters predicts.

However, we emphasize that these conclusions are extremely tentative because of the minimal amount of data available for comparison, the mostly disjoint stellar mass ranges probed by our simulations vs. the observations, and especially the large systematic uncertainties in the observational constraints which are not necessarily captured in the statistical uncertainties on quoted errors: dust masses are derived from infrared luminosities, which depend on dust mass, the dust extinction coefficient, and strongly on the dust temperature. The latter two are highly uncertain in high redshift galaxies and difficult to independently constrain. In addition, our simulations may also miss some real sources of scatter, as we discuss further below. Consequently, the most robust constraints will come from forward modelling of directly observable quantities.

\subsection{Forward-Modeled Observable Quantities: Comparison to Data}

\begin{figure*}
    \centering
    \includegraphics[width=\textwidth]{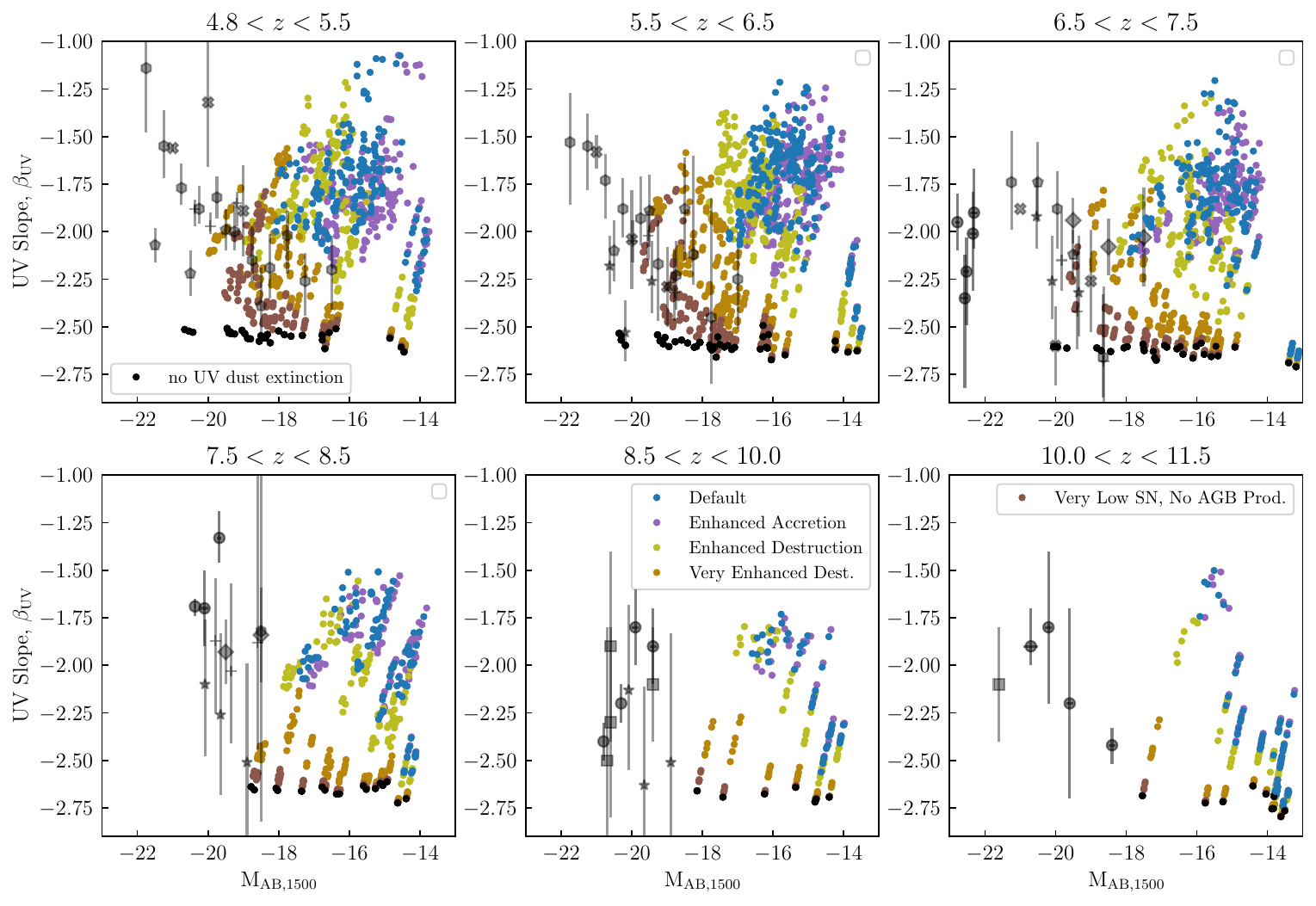}
    \caption{$\beta_{\rm UV}$ as a function of UV AB absolute magnitude. Different panels show data for different bins in redshift, different colors are different dust models, the black points indicate values predicted in the absence of dust, and the grey points are a compilation of observational measurements from the literature: \citet[][plus-signs]{Finkelstein2012}, \citet[][hexagons]{Bouwens2014}, \citet[][diamonds]{Dunlop2013}, \citet[][stars]{Bhatawdekar2021ApJ}, \citet[][filled x]{Wilkins2011}, \citet[][pentagons]{Dunlop2012}, and \citet[][squares]{Wilkins2016} show sample averages of multiple galaxies, while  circles show measurements of inidividual galaxies with JWST from \citet{Roberts-Borsani2022, Naidu2022, Robertson2023}, and \citet{Whitler2023}.}
    \label{fig:beta_UV_M_AB_UV_corr}
\end{figure*}

\begin{figure*}
    \centering
    \includegraphics[width=\textwidth]{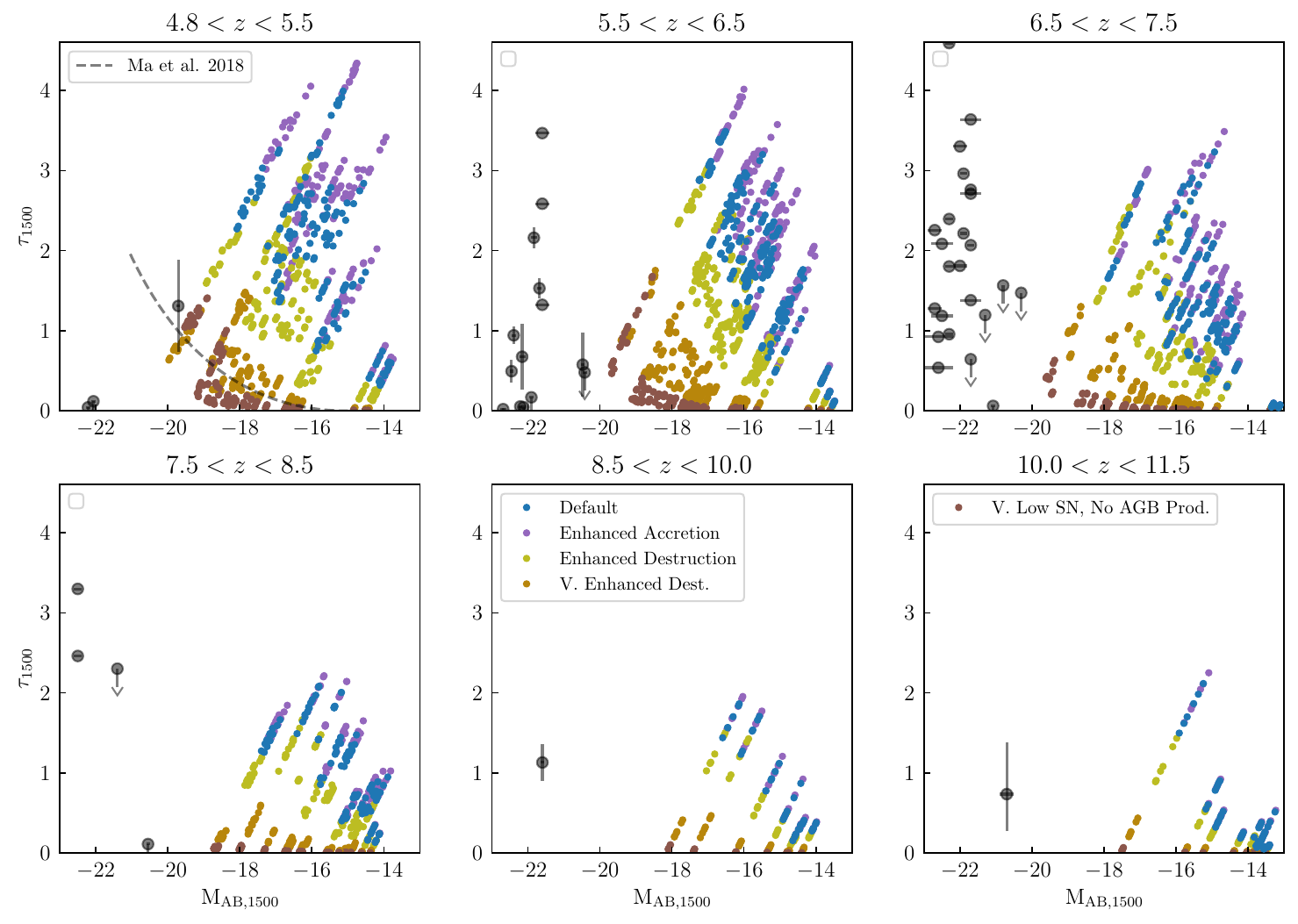}
    \caption{Dust optical depth in the UV vs. absolute UV magnitude. Observational upper constraints from \citet{Schaerer2015}, \citet{Burgarella2020}, \citet{Naidu2022}, and \citet[][with absolute UV magnitudes taken from \citet{Bouwens2022_REBELS}]{Ferrara2022} are shown. In the lowest redshift bin we also show the predictions from \citet{Ma2018} where high resolution galaxy simulations were post-processed with a simpler dust model (see see Section \ref{subsec:discussion_comparison} for discussion).}
    \label{fig:tau1500_M_AB_UV_corr}
\end{figure*}

For the remainder of our analysis we focus on a representative subset of the models discussed previously, each with a consistent color throughout the figures: {\bf Default} (blue), {\bf Enhanced Accretion} (purple), {\bf Enhanced Destruction} (yellow), {\bf Very Enhanced Destruction} (light brown), and {\bf Very Low SN, No AGB Production} (dark brown). 

\subsubsection{Rest-Frame UV Observables: $\rm M_{AB}$, $\beta_{UV}$, $\tau_{1500}$}
\label{subsubsection:UV_obs}

Figure \ref{fig:beta_UV_M_AB_UV_corr} compares the predicted ultraviolet spectral slopes of our simulated galaxies with these dust models to observational data. We also show the predictions of the simulations absent dust extinction in black points. In contrast to the suggestions of Figures \ref{fig:MD_Mstar_corr_acc_dest} and \ref{fig:MD_Mstar_corr_yD}, the models with the lowest dust content -- {\bf Very Enhanced Destruction}, and {\bf Very Low SN, No AGB Production} -- agree best with the data, at all redshifts. It is not clear, however, if either model alone predicts as much scatter at a given luminosity as shown in the data.

In contrast, the more dust-rich models all predict similar $\beta_{\rm UV}$ values which fail to overlap with the observations at any redshift, and exhibit large scatter. This is because they predict very high ISM optical depths, as shown in Figure \ref{fig:tau1500_M_AB_UV_corr}. For dust masses greater than or equal to those predicted by the {\bf Enhanced Destruction} model, the dusty ISM is effectively opaque, so changes in dust content do not impact UV properties significantly. The spread in $\beta_{\rm UV}$ for these models is therefore likely due to the Poisson scatter in number of visible star particles along a given line of sight.  Finally, we note that while the data prefer the dust-poor models, they are inconsistent with entirely dust-free predictions (black points), especially at later times.

\subsubsection{Far-Infrared Observables: IRX-$\beta$ Relation, $L_{\rm IR}$}

\begin{figure*}
    \centering
    \includegraphics[width=\textwidth]{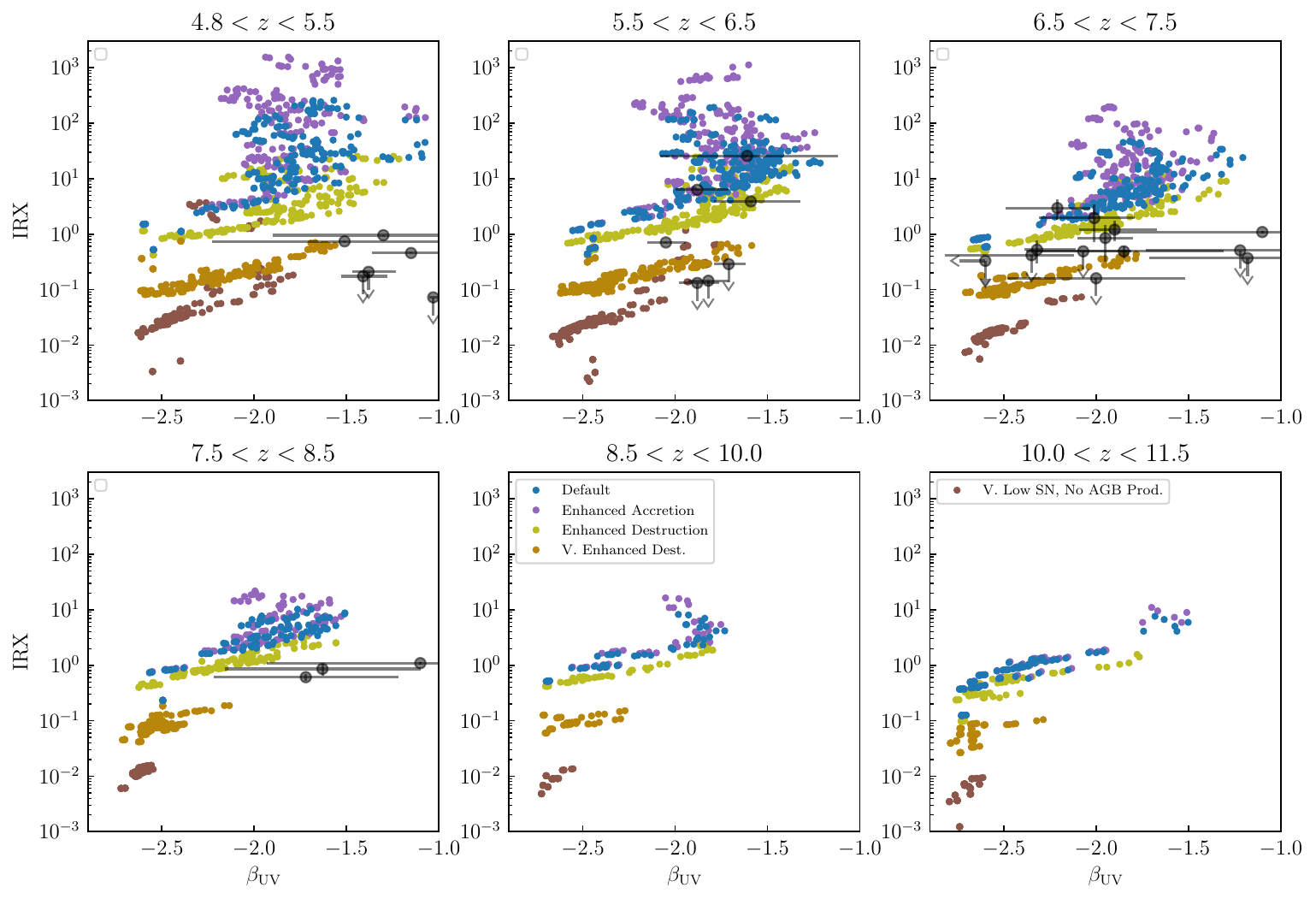}
    \caption{IRX-$\beta_{\rm UV}$ relation. Infrared-Excess (IRX) vs. ultraviolet spectral slope $\beta_{\rm UV}$ for our simulated galaxies in each dust model. Note that a constant dust temperature of $T_D = 40$K was assumed in calculating all infrared luminosities. The colors and redshift bins are identical to Fig.~\ref{fig:beta_UV_M_AB_UV_corr}. Data shown are from \citet{Barisic2017} \citep[which includes data from][]{Capak2015, Pavesi2016}, the compilation from \citet{Hashimoto2019}, and \citet{Bowler2022}.}
    \label{fig:IRX_beta_UV_corr}
\end{figure*}

\begin{figure*}
    \centering
    \includegraphics[width=\textwidth]{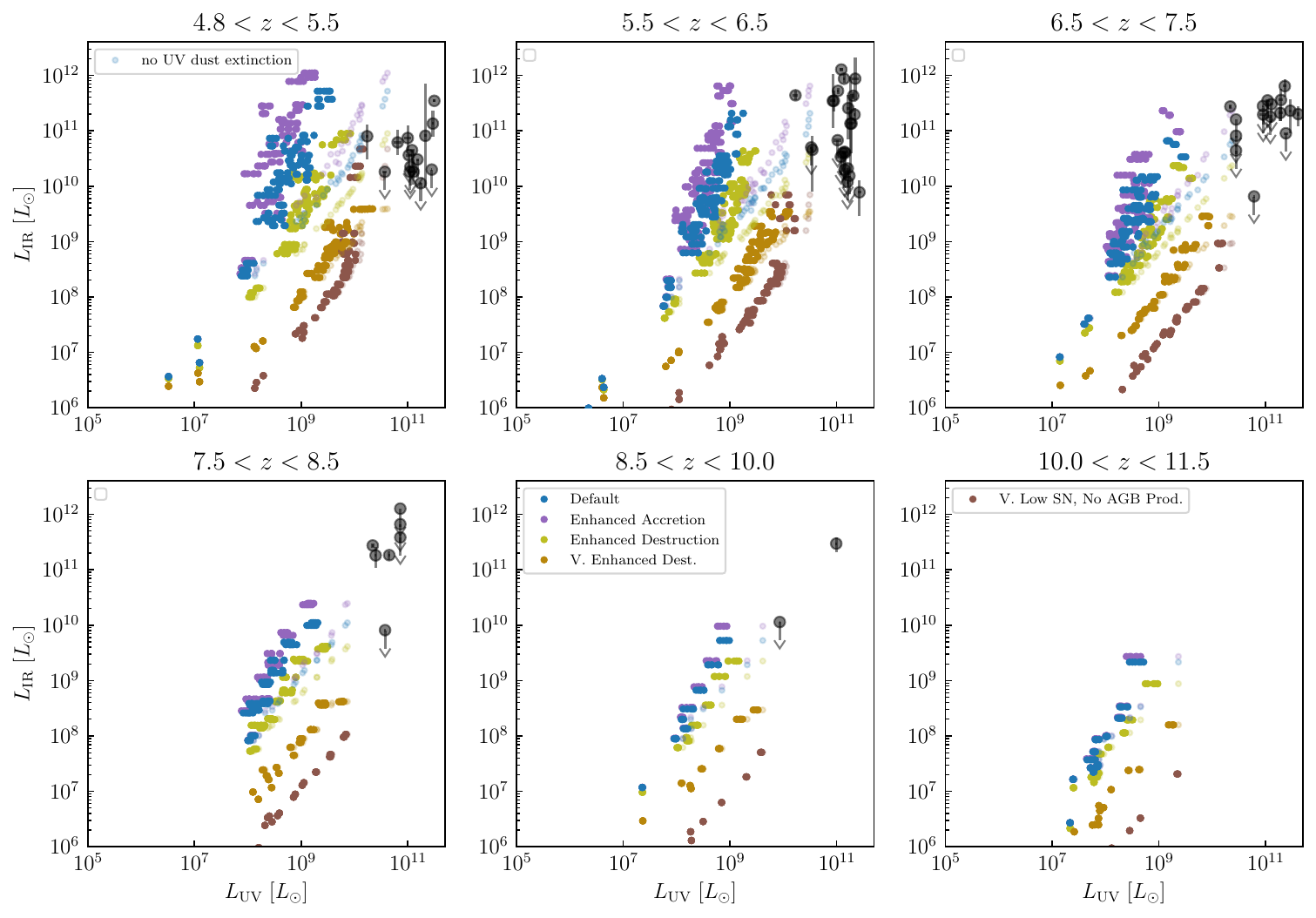}
    \caption{Infrared luminosity vs. UV luminosity. Colors and observational data are the same as Figure \ref{fig:IRX_beta_UV_corr}, with the addition of data from \citet{Burgarella2020}. Additionally, we show predictions without UV dust attenuation in transparent points. These are inconsistent with the simulation, but show the effect of reduced UV opacity with unchanged dust mass.}
    \label{fig:L_IR_L_UV_corr}
\end{figure*}

In Figure ~\ref{fig:IRX_beta_UV_corr} we examine the IRX-$\beta$ relationships predicted by our modeling, compared to observational constraints. Because infrared luminosity depends linearly on dust mass, models with distinctly different dust content are better separated in this parameter space. However, the predictions fail to match the data in two key ways: 1. No one model exhibits as much scatter as the observations, especially in the $5.5 < z < 6.5$ range, and 2. our simulations lack galaxies at low ($\lesssim 1$) IRX and high ($\gtrsim -1.5$) $\beta_{\rm UV}$ which are apparent in the data. 

Some of the reason for these disagreements is illuminated by examining the predictions in $L_{\rm IR}$ vs. $L_{\rm UV}$ space, since these are the numerator and denominator of the IRX, respectively. This is shown in Figure ~\ref{fig:L_IR_L_UV_corr}, along with the same data as in Figure \ref{fig:IRX_beta_UV_corr}. While we predict reasonable $L_{\rm IR}$ especially at late times, all of the galaxies in this observational sample exhibit higher UV luminosities than ours. This helps to explain the lack of low IRX galaxies in our predictions. This suggests that our models are predicting dust masses consistent with observations, but opacities that are too high, in agreement with the interpretations of Figures \ref{fig:MD_Mstar_corr_acc_dest}, \ref{fig:MD_Mstar_corr_yD} and \ref{fig:beta_UV_M_AB_UV_corr}. Indeed, in Figure \ref{fig:L_IR_L_UV_corr} we show the predictions of our simulations without dust attenuation as transparent points, and they are in better agreement with the data. This experiment is nonphysical in that the two luminosities are inconsistently calculated. However, it suggests that real galaxies have similar dust content to our more dust-rich models, but that it is distributed so as to have a much lower effective optical depth.

The inability of any one dust model to reproduce the scatter in observed infrared luminosities could be due to our lack of self-consistently calculated dust temperatures -- for simplicity and given the large modelling uncertainties involved, we assume a constant dust temperature of $T = 40$K for these calculations \citep[see][]{Sommovigo2022}. Since $L_{\rm IR} \propto T_D^4$ at fixed $M_D$, galaxy-galaxy scatter in $T_D$ could significantly enhance the predicted range of IRX. This limitation of our modeling is potentially the primary reason for the low scatter - dust temperatures depend on the radiation field, which, in turn, is sensitive to short time-scale variations in star formation rate. One therefore expects the ISM radiation fields, and consequently dust temperatures, to vary widely from galaxy to galaxy. As well, the lack of very massive galaxies in the limited cosmological volume of our simulation might also mean that we simply aren't sampling galaxies as massive as those in existing observational samples, and this may also be the cause of one or both of these discrepancies to some degree.

It is of course also possible that the dust dynamical quantities assumed as inputs into each model, such as the characteristic time for dust grain growth in the ISM or the supernova destruction efficiency -- vary from galaxy to galaxy due to differences in ISM phase structure and dust content that our model is not sophisticated enough to capture. Infrared observables will therefore require significantly further theoretical efforts to be used as constraints on dust physics at high redshift.

\subsection{Spatial Analysis}
\label{subsection:spatial}

\begin{figure*}
    \centering
    \includegraphics[width=\textwidth]{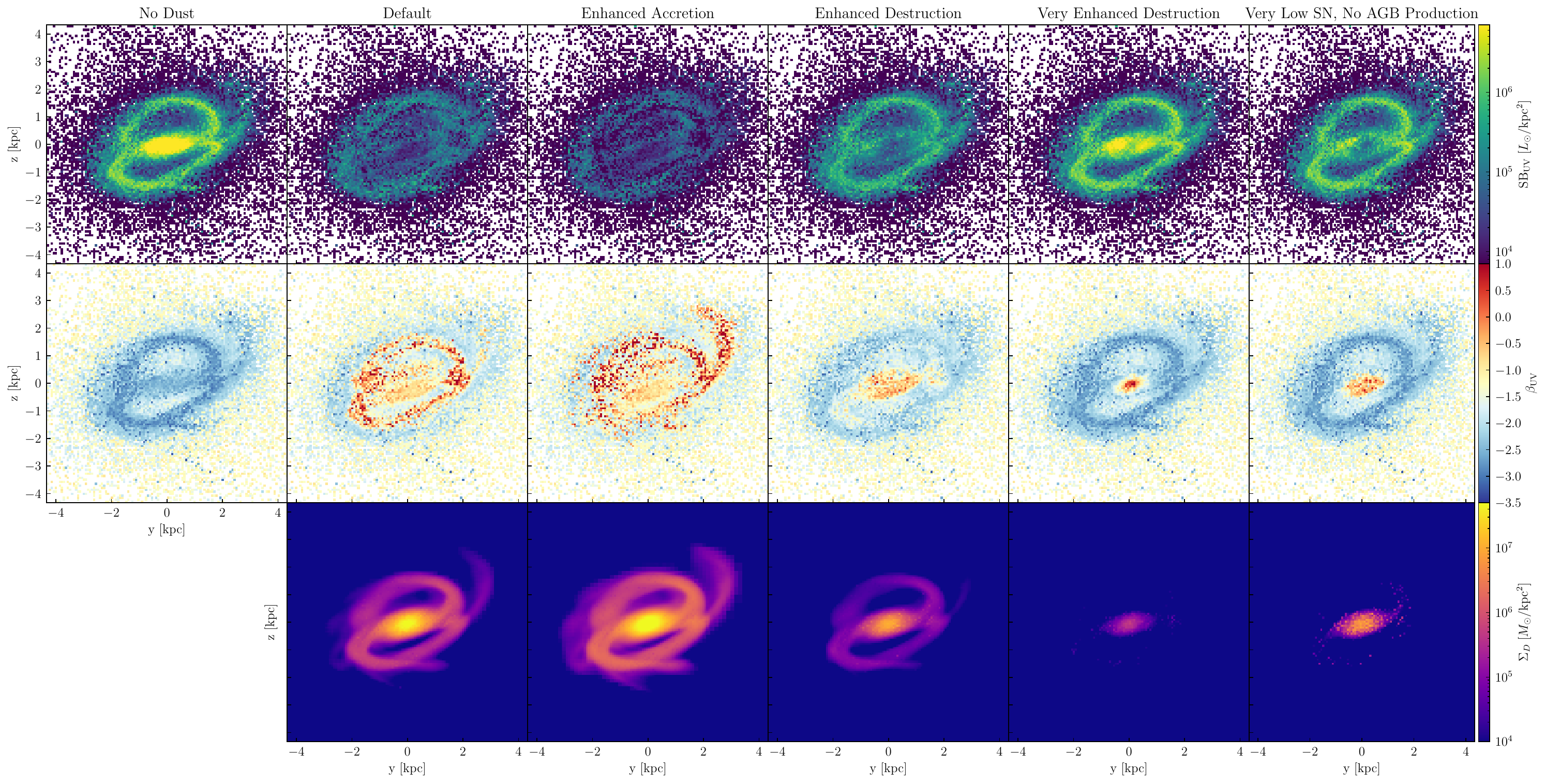}
    \caption{Images of the most massive galaxy in our box along a random line-of-sight at $z=5$ with different dust models. The top row shows the 1500$\angstrom$ UV surface brightness, the middle row shows the spatially resolved UV beta slope (estimated using the 1500$\angstrom$ and 2500$\angstrom$ color) and bottom row shows the column density of dust mass (which is proportional to the IR surface brightness in the optically thin regime). Each column shows the predictions of a different set of dust model parameters, as well as the intrinsic UV emission on the leftmost column. Note that no smoothing has been applied to these images, and the pixelation is the result of the simulation grid.}
    \label{fig:images}
\end{figure*}

\begin{figure*}
    \centering
    \includegraphics[width=\textwidth]{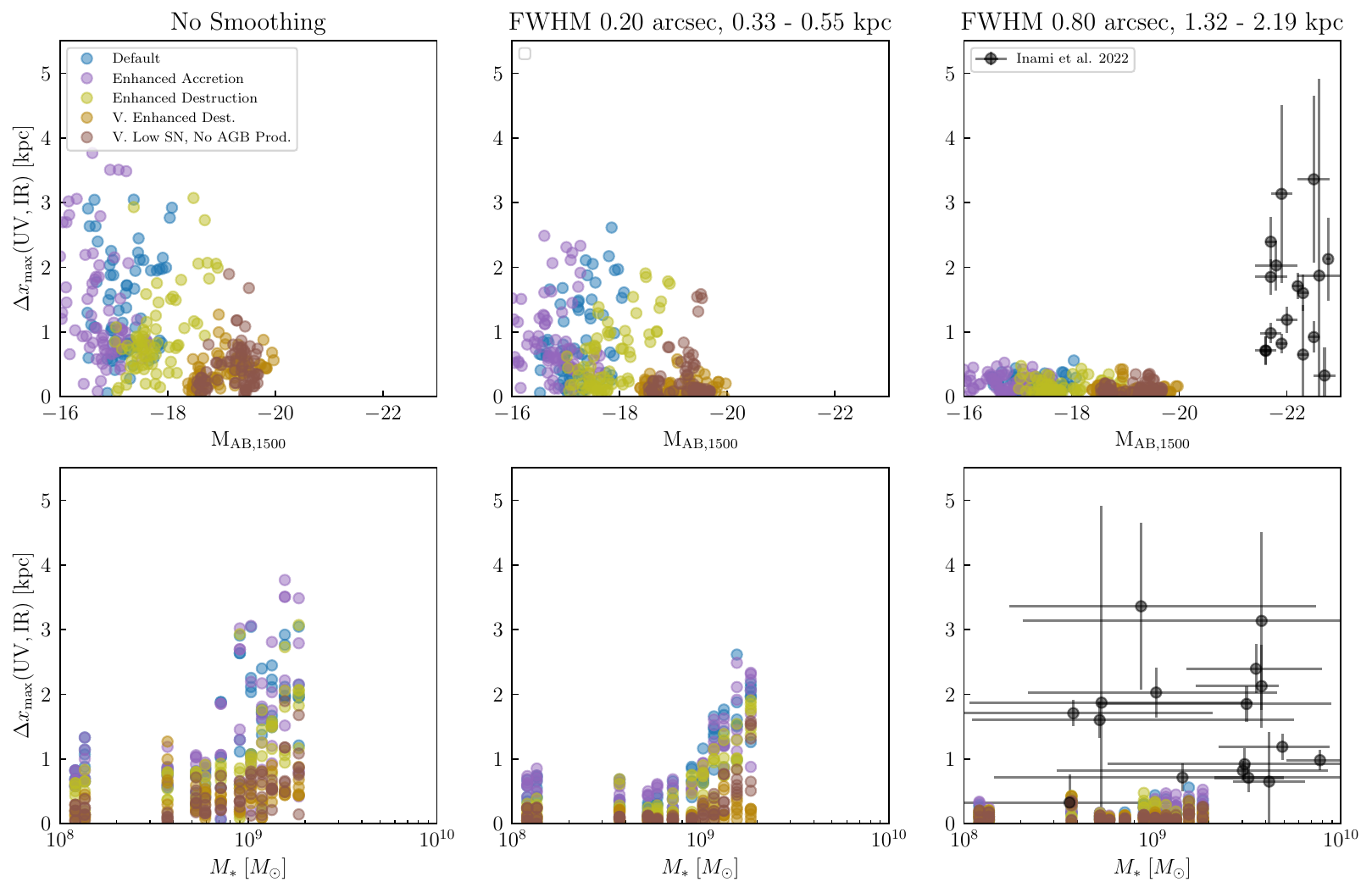}
    \caption{UV and IR peak emission offsets. The projected physical distance between the maximum UV emission (accounting for dust attenuation) and the maximum IR emission (as determined by the dust surface density), as a function of UV absolute magnitude (top row) and stellar mass (bottom row). Each point is one of six lines-of-sight for a each snapshot. Data for the most massive galaxy in our box at $5 < z < 8.5$ is shown. Different colors correspond to different dust models. Each panel shows different levels of smoothing to capture the effect of observational resolution. Data on the right-most plot are from Table 4 of \citet[][with stellar masses from \citet{Bouwens2022_REBELS} and \citet{Schouws2022}]{Inami2022}, whose observations have approximately $0.8$ arcsecond resolution in both the IR and UV \citep{McCracken2012}.}
    \label{fig:offsets}
\end{figure*}

\begin{figure}
    \centering
    \includegraphics[width=\linewidth]{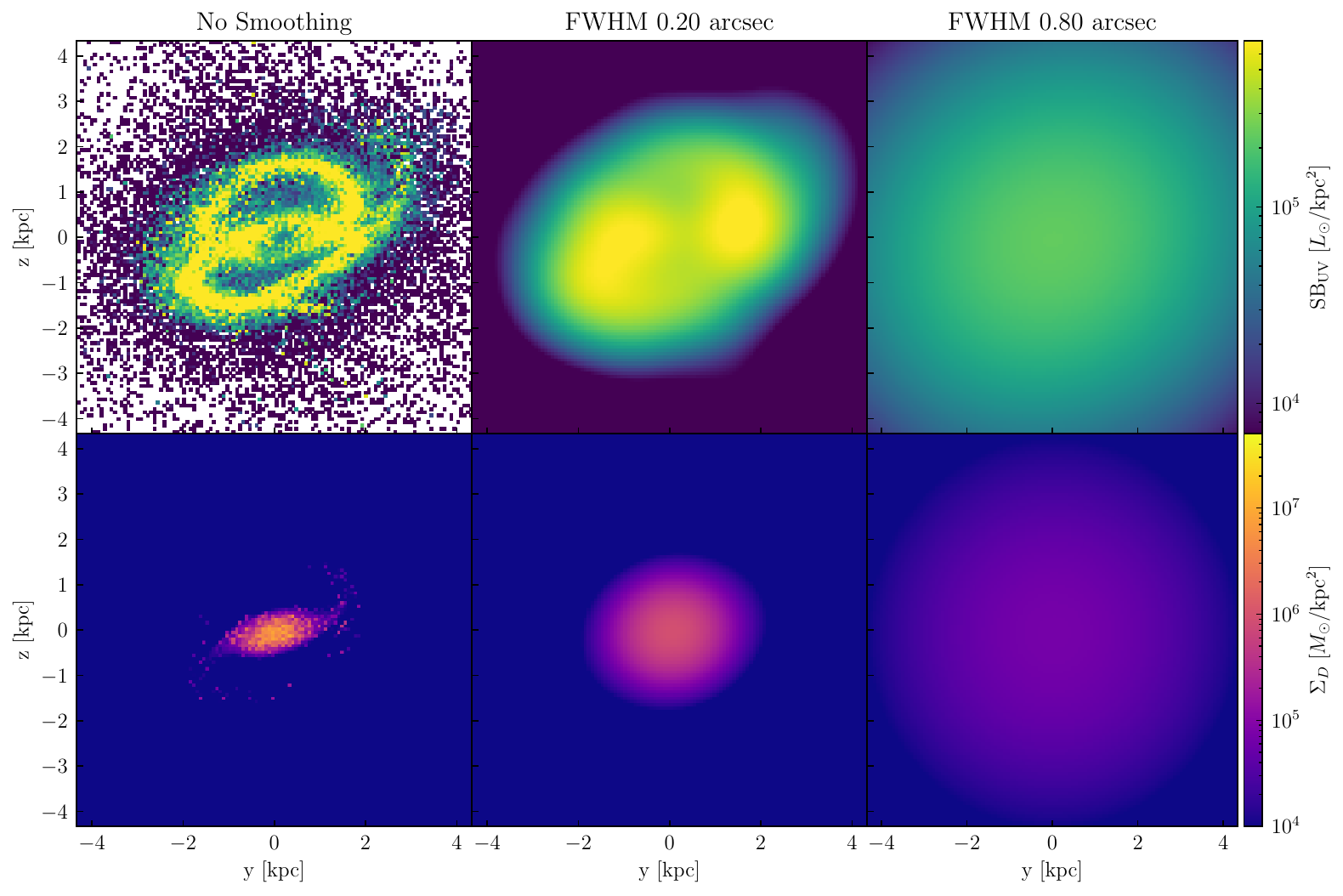}
    \caption{Effect of observation resolution on UV and IR morphologies.}
    \label{fig:smooth}
\end{figure}

Figure \ref{fig:images} shows the predicted UV (1500$\angstrom$) emission, UV color $\beta_{\rm UV}$, and dust column density (which is proportional to the IR emission) for different dust models. Consistent with the results of Section \ref{subsubsection:UV_obs}, all of our dust models predict significant extinction and reddening, but the amount and spatial distribution vary markedly between the different models. The most dust-rich models show large amounts of reddening and extinction throughout the galactic disk, while those with less dust have effects that are more centrally concentrated. However, even those models with the least dust exhibit substantial attenuation and reddening in the center. This is because even the {\bf Very Enhanced Destruction} model predicts dust column densities in excess of $\Sigma_D = 10^5 M_{\odot}/{\rm kpc}^2 \approx 10^{-5} {\rm g}/{\rm cm}^2$ which, for our assumed dust opacity of $\approx 10^5 {\rm cm}^2/{\rm g}$ at 1500$\angstrom$ results in unity optical depth, and $\beta \gtrsim 0$ colors. The increasing severity of extinction at smaller projected galactocentric radii gives the UV emission a ring-like morphology in all but the least dusty model. Color is strongly correlated with IR emission though not perfectly -- for example, both {\rm Default} and {\rm Enhanced Accretion} models exhibit reddest colors in the disk outskirts / tidal tails, while the IR emission is highest in the center. 

All but the {\bf Very Enhanced Destruction} dust model show offsets in the location of maximum UV and IR emission on the order of 1 kpc for the same reason: the regions of highest IR emission are totally opaque to UV light. Note that this is despite the fact that the unattenuated UV light and dust are largely co-spatial. Figure \ref{fig:offsets} shows the measured offsets in projected distance between locations of maximum UV and IR brightness for the same models, at all simulated redshifts, sampling 6 lines-of-sight (positive and negative coordinate axes, which should be random with respect to the galaxy orientation) per snapshot. The top panels show these offsets as a function of absolute UV magnitude, on the bottom panels as a function of stellar mass. The left-most panel of this figure is consistent with the trends noticed in Figure \ref{fig:images}. Larger dust masses result in larger projected regions in which the dust is totally opaque to UV light. The maximum UV emission thereby happens at the larger projected radii where dust becomes optically thin, while the peak IR emission is always in the galactic center. 

The middle and right panels show the results for the same images smoothed by a Gaussian beam of FWHM 0.2 and 0.8 arcseconds, respectively. The physical scale of the smoothing therefore depends on the snapshot redshift, and corresponds to 0.33 (1.32) kpc at $z = 11.4$ and 0.55 (2.19) kpc at $z=5$ for a FWHM of 0.02 (0.08) respectively. The numbers were chosen to approximately match the resolutions of HST observations and ground-based (e.g. ALMA and UltraVISTA) observations, respectively. On the right-most panel we show data from \citet{Inami2022} of UV-bright $z \sim 6-7$ galaxies for which dust continua was observed with ALMA and rest-frame UV (observation-frame near-IR) was observed as part of the UltraVISTA survey, both of which have approximately 0.8 arsec FWHM resolution \citep{McCracken2012}. While these galaxies are clearly much brighter in the UV than ours, some exhibit very large UV-IR offsets that our simulated galaxy fails to exhibit when smoothed appropriately for comparison. 

Indeed, we see that increased smoothing monotonically reduces the peak emission offset. Figure \ref{fig:smooth} demonstrates why: the angular symmetry of both the stellar and dust distributions result in a ring-like morphology of the UV emission at projected galactocentric distances where the dust becomes optically thin. While the offset of UV and IR maxima at infinite observational resolution is approximately the radius of this ring, at smoothing scales comparable to this radius the UV light is maximized in the center, co-spatial with the peak IR emission. The fact that this holds true for all sightlines in all snapshots for every dust model indicates that the UV-IR morphologies are similar in all cases. Peak UV and IR emission are never asymmetrically offset in a way that is preserved with degrading resolution. We have confirmed with a visual inspection of all snapshots that these conclusions are generic to our simulation at all relevant cosmological times. This generality leads us to strongly suspect that it would hold for higher-mass galaxies simulated with CROC physics and our dust model. This is especially the case since more massive galaxies would be expected to have greater dust masses (see Figures 3.5 and 3.6).

\section{Discussion}
\label{discussion}

\subsection{Status of Dust Constraints}

We begin the discussion by assessing the success of our dust modeling efforts compared to current emperical constraints. There is no one model which appears to agree with all of the existing data. While the infrared luminosities (and therefore estimated dust masses) appear to be best reproduced by models with comparatively higher dust content from high production yields and efficient growth, the models with the lowest dust content are in best agreement with $\beta_{\rm UV}$ constraints. We also note that none of our dust models individually reproduces the scatter in $L_{\rm IR}$ seen in observations. This could be do to the limited halo mass range of our simulation sample, the assumption of a fixed dust temperature for all galaxies (which, in reality, must depend on the ISM radiation field and the latter is expected to vary  on short time scales and from galaxy to galaxy),  the simplicity of our dust model, or some combination thereof. Together, these results suggest that while our model is capable of producing dust masses similar to those of real early-universe galaxies, doing so results in UV opacities that are too high. We speculate that this could be due to the spatial distribution of dust relative to stars -- that our galaxies are too uniform compared to the very turbulent ISM of real reionization-era galaxies, which we discuss in the context of our spatially resolved analysis below.

\subsection{Comparison to Similar Theoretical Work}
\label{subsec:discussion_comparison}
\citet{Lewis2023} recently presented results of an investigation with similar aims: they coupled an explicit model for dust very similar to ours to a galaxy formation simulation of cosmological reionization, and use this to predict the dust content and rest-frame UV observables of high redshift galaxies. They present predictions for a single choice of dust model parameters, in which they assume very low dust production yield $y_D = 10^{-3}$ and a much higher ISM growth accretion rate -- they adopt a modestly shorter characteristic timescale (100 vs. 300 Myr) and their expression has an additional factor of $1 / Z \sim 10^3 - 10^4$. Consequently their galaxies transition from production-dominated to accretion-dominated dust content at a lower galaxy stellar mass of roughly $10^6M_{\odot}$ in contrast to $\gtrsim10^7M{\odot}$ in all of our models. 

Their dust masses are therefore most similar to our highest-dust-content models. However, their model predicts much milder UV dust attenuation than ours with similar dust masses, see their Figure 7. We speculate that this is due to differences in resolution: their maximum physical resolution is an order-of-magnitude poorer than ours, at $\sim 1$kpc. Given the observed sizes of high redshift galaxies are $\lesssim 1$kpc \citet{Bouwens2020}, their galaxies cannot possibly be spatially resolved and are therefore likely artificially large. This spreads the same amount of dust over a larger surface area and consequently reduces their predicted optical depths. 

\citet{Graziani2020} also recently conducted simulations of high-redshift galaxies with a coupled dust physics model. Again, they only present predictions for a single set of dust model parameters, which appear to make qualitatively similar predictions between  our {\bf Default} and {\bf Enhanced Accretion} models: their Figure 4 indicates a production yield of $y_D \sim 0.1$, and a transition to accretion-dominated dust at $Z \sim 3\times 10^{-2}Z_{\odot} \sim 6\times10^{-4}$. We note however that they adopt both a much shorter characteristic timescale for ISM accretion of 2 Myr to our 300 Myr, and a somewhat different accretion rate scaling of $dD/dt \propto D Z$ as opposed to our $dD/dt \propto D (f^{\rm dep}Z - D)$. The two expressions tend to the same value at low $D$ (modulo $f^{\rm dep}$ factor which is order-unity), but will differ significantly at higher $D$ -- while ours will tend to zero as $D\to Z$, corresponding physically to all the available metals being locked in dust grains, theirs increases unbounded. This suggests that the plateau in $D/Z$ at $\approx 0.4$ with increasing $Z$ exhibited by their model is due to enhanced destruction rates which regulate the dust content, whereas in our model the transition is set by the $f^{\rm dep}Z-D$ term in the growth rate going to zero.

Moreover, the fact that their model transitions to accretion-dominated at similar $Z$ to ours despite a much higher growth-rate normalization suggests significant differences in the cold gas fractions or thermodynamics of the ISM in our simulations. We also note that their simulation accounts for dust dynamical effects at run time, which our post-processing model cannot. All of this suggests that 1. the predictions of these dust models are sensitive to the implementation details of cooling, star formation, and feedback in the ISM of high redshift galaxies, and/or 2. that the back-reaction of dust dynamics on the ISM might be significant. 

We also consider recent analyses that predict the dust content of high-redshift galaxies with simpler post-processing physical dust models, but with much higher-resolution ($\sim 10$pc) simulations that can more realistically resolve ISM dynamics and phase structure. \citet{Ma2018, Ma2019} predict dust-sensitive quantities from high resolution simulations from the FIRE project \citep{Hopkins2018} of galaxies at $z\ge5$ by assuming a constant $D/Z = 0.4$. While their analysis predict similar dust masses to our more dust-rich models -- see the dashed grey line in Figure \ref{fig:MD_Mstar_corr_acc_dest}, their predicted effective optical depths are most consistent with our least dusty models (see Figure \ref{fig:tau1500_M_AB_UV_corr}). Interestingly, this is also the case with the results from a similar study \citep{Mushtaq2023} using the FirstLight simulation suite \citep{Ceverino2017} and an identical dust post-processing model -- see Figures 1 and 5 in \citet{Mushtaq2023}. 

FIRE and FirstLight are different galaxy formation simulations of similarly high resolution, both significantly higher than ours. Consequently, they better capture the effects of feedback on the high-redshift interstellar medium, resulting in a more turbulent, porous gas distribution which we speculate has a broader column density distribution than ours -- see Figure 4 of \citet{Ma2019} and Figure 1 of \citet{Ceverino2021}, both of which exhibit large gas column density fluctuations on scales smaller than our 100 pc resolution. This results in lower effective optical depths at a given dust mass because there exist low-density column channels through which UV radiation can escape that are lacking in our simulation. 

We have conducted an analysis of the predicted UV, IR and UV color morphology of the most massive galaxy in our simulation under the assumption of different dust model parameters. We find that all models predict significant dust attenuation in the central region of the galaxy, resulting in red $\beta \gtrsim -1$ colors and in all but the model with the least dust ring-like morphology for the UV emission. This is because the dust contents predicted by our models are generally optically thick in a region that is approximately symmetrical about the galactic center, so the UV emission is dominated by the smallest radii at which dust becomes optically thin. Color is also strongly correlated with dust column, which we use as a proxy for IR emission.

Since IR emission peaks in the center of the galaxy, there are $\sim$ kpc-scale offsets between the points of maximal UV and IR surface brightness when ``observed'' with infinite resolution, but degrading image resolution on scales similar to existing observational capabilities causes the UV emission to peak in the center due to its symmetric distribution, resulting in no offset between peak brightness in UV and IR. While existing observations only probe galaxies brighter in the UV than the most massive in our sample, they do exhibit much larger offsets that are suggestive of more complicated morphologies than the ones predicted by our modeling efforts, see Figure 2 of \citet{Bowler2022} and Figure 7 of \citet{Inami2022}. Indeed, Figure 3 of \citet{Bowler2022} displays UV color gradients much less symmetric than any of those predicted by our dust modelling.

We note that the analysis of galaxy-averaged observable properties would lead us to expect that the distributions of UV and IR emission predicted by our models would be overly smooth and symmetrical, given our inability to simultaneously match observed dust masses and optical depths. We interpreted this as evidence that our simulations fail to reproduce a sufficiently dynamic ISM and consequently the full distribution of dust column densities, the lower tail of which could allow for significantly enhanced UV emission without decreased dust mass. The results of this spatially resolved analysis provide evidence in favor of this interpretation given the inability of our modelling to reproduce the asymmetric morphologies seen in data of galaxies with similar stellar masses.

Simulation resolution and feedback prescription are the two most important numerical components of a fluid-dynamical galaxy formation model for determining the structure and dynamics of the ISM, and therefore one or both of these is likely implicated in our modelling failures. At a spatial resolution of 100pc, our simulations do not resolve the disk scale-height and therefore cannot capture fully three-dimensional phenomena that characterize the ISM phase structure like molecular clouds and supernova feedback ``super-bubbles''. As a consequence, the delayed cooling feedback prescription utilized in CROC appears to be incapable of driving large-scale galactic winds -- we have watched movies of the tracer particles used in this analysis and they are never removed from the galaxy ISM, indicating a negligible mass flux from the ISM into the circumgalactic medium. This is in stark contrast to most other modern galaxy formation models in which galaxies of the relevant mass range drive strongly mass-loaded winds, especially at early cosmological times \citep[e.g.][]{Muratov2015, Pandya2021}. A feedback prescription that successfully launches winds would reduce the gas mass and therefore dust mass in our galaxies, possibly reducing the high opacities of our most dust-rich models. These winds might also carve out low column density sight-lines with minimal dust extinction.

\citet{Ma2018}, \citet{Ma2019}, and \citet{Liang2021} thoroughly explored the UV-to-IR observable properties of reionization era galaxies predicted by the FIRE-2 simulations, which are significantly higher resolution than ours ($\sim 10$pc) and have been demonstrated to drive galactic winds. While they do not explicitly quantify any offsets between predicted UV and IR emission in their simulations, we see suggestions from their analysis of the dynamic, asymmetric ISM seen in observations and lacking in our simulations. Figure 4 of \citet{Ma2019} shows images of UV light and dust column density for two of their simulated galaxies, which both display a much more disturbed morphology than anything we find in our analysis. Close inspection reveals that the regions of brightest UV surface brightness correspond to holes in the dust surface density which appear to be blown out by strong feedback. However, we note that the spatial offsets between peak UV and IR emission do not visually appear to be much larger than 1 kpc, but firm conclusions cannot be drawn from images of just two galaxies each at a single snapshot. Figure 12 of \citet{Liang2021} does explicitly show a galaxy with $>1$kpc offset between maximum UV and IR surface brightness, due to a highly perturbed and asymmetric distribution of gas with respect to stars (though we note that this galaxy is significantly more massive than those in our analysis). They also find that the effective UV optical depth does not correlate with dust mass at all at high redshift $z=6$ because of large variations in the star-dust geometry predicted by their simulations. All of this suggests that higher resolution simulations with a feedback model that drives galactic winds may be better able to match the asymmetric UV/IR morphologies seen in observations.

The SERRA project is another suite of high-resolution cosmological simulations of galaxies at $z>6$. These simulations are higher resolution than ours by about a factor of 3 with minimum cell sizes of $\sim 30$pc, and consequently have different star formation and feedback prescriptions, more similar to those in FIRE-2 \citep{Behrens2018, Pallottini2022}. In contrast to our work and similar to FIRE, they find a clumpy morphology for both stars and dust, which in some cases leads to spatial offsets \citep{Pallottini2022}. They also find that this clumpiness results in low effective optical depths due to dust, although star-forming regions can locally exhibit very high extinctions \citep{Behrens2018}. It is interesting to note that Figure 4 of \citet{Behrens2018} does appear to exhibit a ring-like morphology in the galaxy's central UV emission, suggesting this effect might persist to higher-resolution simulations. Nonetheless, the relative UV and IR properties of these galaxies are strongly influenced by the presence of dusty, star forming clumps which our simulations could not resolve, suggesting resolution is a main issue for our theoretical predictions.

Our results therefore provide strong motivation for the development of dust models such as the ones presented here in higher-resolution simulations of galaxy formation with more realistic feedback resulting in a manifestly multiphase ISM, as this appears to be essential to capturing the effects to which observations are most sensitive. 

\section{Conclusion}
\label{conclusion}
We apply the dust post-processing model described in \citet{Esmerian2022pub} to a suite of 11 simulated galaxies from the CROC project. We explore 9 different sets of dust parameters and quantify the effect of their variation on the dust content of high redshift galaxies. We then forward model observable properties of high-redshift galaxies and compare to existing data. Our conclusions are the following:
\begin{itemize}
    \item Comparing our simulated galaxies to a compilation of recent constraints on the metallicities of reionization-era systems, we find general agreement, although CROC might slightly under-predict metallicity at a given stellar mass.
    \item We vary dust model parameters governing the rate of grain growth due to accretion in the ISM, the efficiency of grain destruction in supernova remnants, and the dust yields of production sources (supernova and AGB star winds), to determine their impact on the predicted dust contents of high-redshift galaxies. We qualitatively validate the results of \citet{Esmerian2022pub}, in which we reproduced a well-established behavior of these dust models \citep[see][for a review]{Hirashita2013}: the dust content of galaxies is set at early times/low metallicities primarily by the assumed production yields, while at higher metallicities/late times it is set by the competition between accretion and destruction, normalized by the initial condition set by production yields. The transition occurs around $Z \sim 2-4 \times 10^{-4} = 1 - 2\times 10^{-2} Z_{\odot}$, with some dependence on assumed model parameters.
    \item However, we observe significant scatter between galaxies at a constant metallicity, especially at late times/higher metallicities for models in which growth via accretion becomes efficient. This indicates the existence of important secondary dependencies beyond metallicity that determine the dust content of galaxies, which is not captured by typical one-zone models \citep[e.g.][]{Feldmann2015}. We speculate that this is driven by some combination of star formation history and ISM phase structure dependence, as is evidenced by the particularly aggressive growth via accretion in the most massive galaxy compared to the other galaxies in our sample.
    \item We compare our total predicted dust masses as a function of stellar mass to observational constraints in the literature, and while our limited simulation size fails to sample galaxies as massive as most of those with observational constraints, where there is overlap we find our most dust-rich models -- {\bf Default} and {\bf Enhanced Accretion} -- appear to predict scaling relations consistent with current data. This suggests that the data prefer models in which production yields are high and ISM grain growth is efficient at high masses. The data also appear to exhibit larger scatter at a given stellar mass than predicted by any one of our models, but due to both large systematic uncertainties in the dust mass observational constraints and the disjoint range of stellar masses probed by our simulations vs.\ the observations, these conclusions are tentative. Nevertheless, it is easy to imagine several additional sources of scatter that are missed in our simulations and post-processing, such as dependence of the dust temperature on the local radiation field or deficiency of the stellar feedback model.
    \item We forward model directly observable galaxy properties from our simulations to make more direct comparison to data, and find that we are unable to simultaneously match existing observational constraints with any one model. Specifically, the models which best match the observed spectral slope in the UV, $\beta_{\rm UV}$, are the models with least dust content due to either low production yields or very high destruction rates. However, these models fail to predict sufficiently high infrared luminosities. Those that do predict IR luminosities consistent with observations have far too much dust extinction and thereby fail to agree with $\beta_{\rm UV}$ constraints. Finally, we note that no one of our models appears to predict as much scatter in these observable quantities as the data exhibit.
    \item We speculate that these deficiencies are due to issues with the spatial distribution of dust relative to stars in our simulations, which may be overly smooth. To assess this hypothesis, we compare our simulations to spatially resolved observations of rest-frame UV emission and dust continuum \citep{Inami2022}, between which some galaxies show large spatial offsets, indicative of a highly dynamic ISM. We compare data from galaxies of similar estimated stellar mass to our most massive system, and find that all of our models fail to predict offsets as large as observed, lending support to the idea that our galaxies fail to capture the dynamic complexity of the high-redshift ISM, which is necessary to reproduce observations.
\end{itemize}


\acknowledgments
This manuscript has been co-authored by Fermi Research Alliance, LLC under Contract No. DE-AC02-07CH11359 with the U.S. Department of Energy, Office of Science, Office of High Energy Physics. This work used resources of the Argonne Leadership Computing Facility, which is a DOE Office of Science User Facility supported under Contract DE-AC02-06CH11357. An award of computer time was provided by the Innovative and Novel Computational Impact on Theory and Experiment (INCITE) program. This research is also part of the Blue Waters sustained-petascale computing project, which is supported by the National Science Foundation (awards OCI-0725070 and ACI-1238993) and the state of Illinois. Blue Waters is a joint effort of the University of Illinois at Urbana-Champaign and its National Center for Supercomputing Applications. This manuscript is based upon work that is supported by the Visiting Scholars Award Program of the Universities Research Association. This work was completed in part with resources provided by the University of Chicago’s Research Computing Center. Our analysis made use of the following publicly available software packages: Matplotlib \citep{Matplotlib}, SciPy \citep{SciPy}, NumPy \citep{NumPy}, COLOSSUS \citep{COLOSSUS}, and yt \citep{Turk2011}. This manuscript is presented as part of a thesis to the Department of Astronomy and Astrophysics, The University of Chicago, in partial fulfillment of the requirements for the Ph.D. degree.

\appendix

\section{$L_{\rm IR}$ Correction}
\label{appendix:L_IR}

\begin{figure}
    \centering
    \includegraphics[height=0.5\textheight]{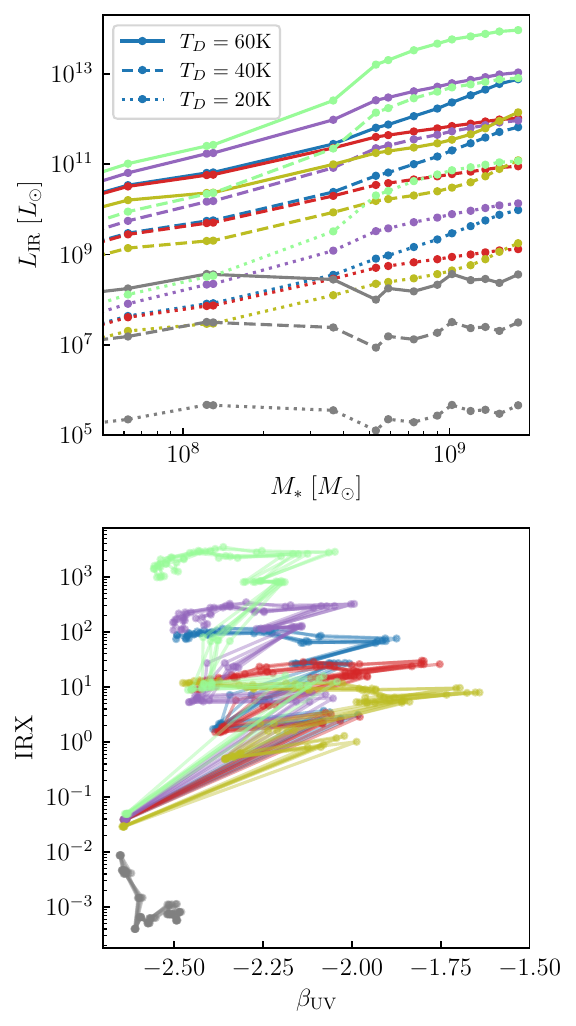}
    \caption{Reproduction of Figure 19 in \citet{Esmerian2022pub}, with corrected calculation of IR luminosities.}
    \label{fig:L_IR_IRX_beta}
\end{figure}

Figure \ref{appendix:L_IR} shows the infrared luminsoity of the most massive galaxy in our simulation as a function of stellar mass (top panel), and the IRX-$\beta_{\rm UV}$ relationship for the same galaxy (bottom panel). Different colors represent different dust model parameter choices as in \citet{Esmerian2022pub}. In the top panel, different line-styles indicate different choices for the dust temperature (dotted: $T_{\rm D} = 20$K, dashed: $T_{\rm D} = 40$K, solid: $T_{\rm D} = 60$K), while the bottom panel only shows the predictions for $T_{\rm D} = 40$K. This figure is identical to Figure 19 in \citet{Esmerian2022pub}, but with the corrected calculation of infrared luminoisities as described in Section \ref{methods}.

\bibliography{bibliography}
\end{document}